# Topological Materials for Near-Field Radiative Heat Transfer

Azadeh Didari-Bader, [1, †], Seonyeong Kim[1,2, ††], Heejin Choi[1,2], Sunae Seo[2], Piyali Biswas[1,*], Heejeong Jeong[4, ‡,*], and Chang-Won Lee[1,3,*]

[1]*Institute of Advanced Optics and Photonics, Hanbat National University, Daejeon, Republic of Korea*

[2]*Department of Physics, Sejong University, Seoul, Republic of Korea*

[3]*Department of Applied Optics, School of Basic Sciences, Hanbat National University, Daejeon, Republic of Korea*

[4]*Department of Physics, Faculty of Science, University of Malaya, Kuala Lumpur, Malaysia*

[*]Corresponding authors Emails: pbiswas@hanbat.ac.kr, jhj413@gmail.com, cwlee42@hanbat.ac.kr

[†]Also at Fowler School of Engineering, Chapman University, California, USA

[††]Also at Paul Scherrer Institut, École polytechnique fédérale de Lausanne, Switzerland

[‡]Also at PASQAL Korea, 311 Gangnam-daero, Seocho-gu, Seoul 06628, Republic of Korea

## ABSTRACT

Topological materials provide a platform that utilizes the geometric characteristics of structured materials to control the flow of waves, enabling unidirectional and protected transmission that is immune to defects or impurities. The topologically designed photonic materials can carry quantum states and electromagnetic energy, benefiting nanolasers or quantum photonic systems. This article reviews recent advances in the topological applications of photonic materials for radiative heat transfer, especially in the near field. When the separation distance between media is considerably smaller than the thermal wavelength, the heat transfer exhibits super-Planckian behavior that surpasses Planck's blackbody predictions. Near-field thermal radiation in subwavelength systems supporting surface modes has various applications, including nanoscale thermal management and energy conversion. Photonic materials and structures that support topological surface states show immense potential for enhancing or suppressing near-field thermal radiation. We present various topological effects, such as periodic and quasi-periodic nanoparticle arrays, Dirac and Weyl semimetal-based materials, structures with broken global symmetries, and other topological insulators, on near-field heat transfer. Also, the possibility of realizing near-field thermal radiation in such topological materials for alternative thermal management and heat flux guiding in nano-scale systems is discussed based on the existing technology.

## 1. Introduction

Topology, the mathematics of conserving properties under continuous deformations (homeomorphism), has emerged as a major growing field in recent years with applications to various fields of science [1]. The implementation of topology introduces a range of new possibilities in various branches of physics, chemistry, and material sciences [2–11]. The concept of topological photonics has been rooted in the idea of topological phases of matter in the context of condensed matter. This field was inspired by the discovery of topological insulators, which are insulators in bulk; however, on the surface, they have conductive edges. After the discovery of the integer quantum Hall effect in 1980, various other ideas that subsequently shaped the formation of these insulators were investigated. Although electrons and photons follow different quantum mechanical wave equations, solutions to corresponding eigenvalues (or generalized eigenvalues) are generally anticipated depending on the corresponding self-adjoint operators with structure symmetries in space–time. In the domain of photonics, the topological phases of matter and invariant quantities, that define an object's geometric properties globally, are observed by exploring various carefully engineered photonic structures.

*Topological photonic band theory:* Photonic crystals are the most common platform to explore the topological features, as they exhibit the photonic band structure with discrete levels of eigenstates forming bands separated by band gaps. Eventually, the topology of the bulk originates from the winding of eigenstates in the momentum space that readily acquires a geometric phase, termed Berry's phase [2],

$$\gamma = \oint \boldsymbol{\mathcal{A}}_n(\boldsymbol{k}).d\boldsymbol{k} \qquad (1)$$

where, $\boldsymbol{\mathcal{A}}_n(\boldsymbol{k}) = i\langle u_{n,\boldsymbol{k}}|\boldsymbol{\nabla}_{\boldsymbol{k}}|u_{n,\boldsymbol{k}}\rangle$ is the Berry connection with $u_{n,\boldsymbol{k}}$ the periodic part of the Bloch state of *n*th band and momentum vector $\boldsymbol{k}$. Moreover, the band topology is characterized by a topological invariant which is zero for topologically equivalent or trivial bands and nonzero for topologically nontrivial bands. Suppose the wave functions associated with a particular energy band can undergo a smooth transformation into the wave functions of another band under an adiabatic translation over the momentum space (the whole first Brillouin zone). In that case, those two bands are defined to be topologically trivial where the band gap remains open. Conversely, suppose the wave functions of the two bands under adiabatic translation can only be interconverted into one another with an associated band gap closing and reopening. In that case, there is a discontinuity in band dispersion that readily changes the bulk topology termed as nontrivial. In general, the geometric phases acquired by the wave functions are quantified by the topological invariant Chern numbers (in the case of 2D and 3D geometries) [2],

$$C_n = \frac{1}{2\pi}\int d^2\boldsymbol{k}\boldsymbol{\Omega}_n(k_x, k_y) \qquad (2)$$

which is estimated over the whole first Brillouin zone where $\boldsymbol{\Omega}_n(\boldsymbol{k}) = \boldsymbol{\nabla}_{\boldsymbol{k}} \times \boldsymbol{\mathcal{A}}_n(\boldsymbol{k})$ is the Berry curvature of the *n*th band. It describes the global properties of the wave functions and hence defines the gapped systems' topology. However, for 1D geometries, Chern numbers cannot be estimated due to the lack of a closed boundary, and the Berry phase (given by Eq. (1) in 1D), known as the Zak phase [12], acquired by the Bloch states acts as a topological invariant owing to its quantized nature. Furthermore, establishing a boundary between two topologically distinct geometries with varying invariant quantities facilitates topology-protected localized states of light at the edge/interface following the bulk-edge correspondence theory. Moreover, such a photonic stream gains unidirectionality owing to the symmetry of the photonic structure. It was first demonstrated in 2009 exhibiting a topologically protected mode even with the existence of defects and disorders, such as electron waves [8]. The robustness of such topological edge states comes from the nontrivial bulk topology which does not change under perturbations. It causes lossless photon energy flow or photonic quantum state delivery that is unconstrained by the shape of the edge and immune to defects and disorders [13].

Recent progress in topological photonics has introduced a range of photonic materials that exhibit various topological phases based on their dimensionality. 2D photonic crystals are the most common with three types of major topological phases – a) Quantum Hall phase which is exhibited by 2D gyromagnetic photonic crystals [14], Floquet-like waveguide array [7], dynamic modulation [15], graphene [16] etc. where the presence of magnetic field breaks the time-reversal symmetry and introduces nonzero Chern numbers; b) Quantum spin Hall phase shown by 2D interfaces between vacuum/metal, dielectric/metal, anisotropic media, and 2D photonic crystals with crystalline symmetries where photons have spin-like quantity to achieve spin-locked features of boundary modes [2,17]; and c) Quantum valley Hall phase in various time-reversal symmetric photonic microstructures with broken inversion symmetry to open a gap at Dirac point located at the valley [2,17]. Besides, Floquet topological phases are observed in photonic structures with periodic temporal modulation similar to spatial modulations in photonic crystals [2]. In 3D, one may observe two major types of topological phases – the 3D gapped phase and the 3D gapless phase [17]. The gapped topological phases in 3D are extensions of all the 2D topological phases stated above with edge states confined in all three spatial dimensions. However, 3D gapless phases have an entirely different origin called Weyl points which has no 2D counterpart. Photonic crystals, bulk materials, metamaterials, etc. may exhibit nontrivial bulk topology around the Weyl degeneracies (a pair of Weyl points with opposite topological charges) and are often called Weyl semimetals [17,18]. It is to be mentioned here that the Weyl points are fundamentally different from the 2D Dirac points and are not protected by global symmetries, thus they cannot be created or annihilated by symmetry-breaking phenomena. In this context, graphene is often considered 2D Dirac semimetal owing to the zero density of states at the Dirac point. Lately, 1D topological phases have gained special attention in photonic crystals and other 1D topological models. Zak phase, which is the Berry phase in 1D, also has quantized values – 0 for trivial, and π for nontrivial topology [12]. The transition of the Zak phase is possible by breaking either inversion symmetry in 1D photonic crystals or chiral symmetry in the case of special 1D topological chains.

*Near-field Radiative heat transfer:* In recent years, with the emergence of topological photonic materials, interest has been focused on exploring the unique properties of topologically protected materials in the context of "fluctuational electrodynamics". The fluctuational electrodynamics is described based on the random motion of atoms due to the increase in their thermal energy or solely due to their quantum fluctuations. A general theory describing the relationship between irreversible processes (dissipation) and noise (fluctuations) which is called fluctuation–dissipation theorem (FDT) was formulated by Callen and his co-workers in 1951 and has been available since then [19–22]. However, Leontovich and Rytov were the first to solve the problem by describing the spectral description of fluctuations of electromagnetic fields and applying thermal radiation in a quasi-static limit [23,24]. Later, in 1953, Rytov derived a more general solution with frequency dependence between correlation and thermal fluctuation [25–27]. He studied the heat transfer in a semi-infinite absorbing body with temperature $T$ and with a vacuum gap from a virtually perfect mirror having zero temperature. In his work, he used Leontovich's approximate boundary conditions to describe the mirror. He considered random thermally excited electromagnetic fields as an excitation source. Later, in 1971, Polder and Hove [28] developed a general formalism with the aid of FDT for the radiative heat transfer between closely spaced macroscopic semi-infinite media with arbitrary dispersive and absorptive dielectric properties. These media have different temperatures and are separated from each other by a vacuum gap of size $d$. Their work differed from Rytov's work [25] in two respects. First, they considered random currents as the source of excitation rather than the random fields considered by Rytov [25]. Second, they studied identical semi-infinite bodies at different temperatures using an exact boundary condition. Their work simplified the heat transfer analysis and proved to be more accurate owing to the exact boundary conditions used.

Extensions of the Polder and Hove formalism and FDT have since been extensively studied in the context of radiative heat transfer in the near field. In the far field, the classical regime of radiative heat transfer, which follows Planck's blackbody predictions, is valid [29]. The foregoing is applicable provided that the media are separated from each other by distances that are considerably larger than the thermal wavelength in the surrounding medium, predicted by Wien's law [30]. When the separation distance among the media is comparable or smaller than the thermal wavelength, Planck's predictions are no longer valid, and the super-Planckian regime of heat transfer is in place [31].

To formulate the FDT and establish its relationship with the near-field thermal radiation, the thermal stochastic Maxwell equations, i.e., Eqs. (3) and (4), are considered:

$$\nabla \times \boldsymbol{E}(\boldsymbol{r},\omega) i\omega\mu_0 \boldsymbol{H}(\boldsymbol{r},\omega) \quad (3)$$

$$\nabla \times \boldsymbol{H}(\boldsymbol{r},\omega) = -i\omega\varepsilon(\omega)\varepsilon_0 \boldsymbol{E}(\boldsymbol{r},\omega) + \boldsymbol{J}^{fl}(\boldsymbol{r},\omega) \quad (4)$$

On the right-hand side of Eq. (4), the fluctuating current density, $\boldsymbol{J}^{fl}$, is incorporated into Ampere's law for non-magnetic materials to represent the electric dipole oscillations.

The fluctuating current is fully described and dependent on the first two moments. The first moment is the ensemble average; it is zero and indicates that the mean thermally radiated electric and magnetic fields are also zero. The second moment is the ensemble average of the spatial correlation function of fluctuating currents which is given by the FDT [32] and shown in Eq. (5):

$$\langle \boldsymbol{J}_{\alpha}^{fl}(\boldsymbol{r}',\omega)\boldsymbol{J}_{\beta}^{fl*}(\boldsymbol{r}'',\omega')\rangle = \frac{4\omega\varepsilon_0\varepsilon''(\omega)}{\pi}\,\boldsymbol{\Theta}(\omega,T)\delta_{\alpha\beta}\delta(\boldsymbol{r}'-\boldsymbol{r}'')\delta(\omega-\omega') \qquad (5)$$

where <> indicates the ensemble average; * represents the complex conjugate; $\alpha$ and $\beta$ are orthogonal components depicting the polarisation state of the fluctuating currents; $\delta_{\alpha\beta}$ is the Kronecker function; and $\delta(\omega-\omega')$ $(\boldsymbol{r}'-\boldsymbol{r}'')$ are the Dirac functions. In addition, the mean energy of an electromagnetic state at frequency *ω* and temperature *T* is given by $\boldsymbol{\Theta}(\omega,T)$:

$$\boldsymbol{\Theta}(\omega,T) = \frac{\hbar\omega}{exp\left(\frac{\hbar\omega}{K_B T}\right)-1} + \frac{\hbar\omega}{2} \qquad (6)$$

where the last term accounts for zero-point energy and does not affect the net radiative heat transfer. The volume integral expressions for the electric and magnetic fields in terms of fluctuating currents observed at location $\boldsymbol{r}$ due to the fluctuating current sources located at $\boldsymbol{r}'$ within volume *V* are presented in Eqs. (7) and (8), respectively:

$$\boldsymbol{E}(\boldsymbol{r},\omega) = i\omega\mu_0\int_V \overline{\overline{\boldsymbol{G}}}^E(\boldsymbol{r},\boldsymbol{r}',\omega).\boldsymbol{J}^{fl}(\boldsymbol{r}',\omega)d^3\boldsymbol{r}' \qquad (7)$$

$$\boldsymbol{H}(\boldsymbol{r},\omega) = \int_V \overline{\overline{\boldsymbol{G}}}^H(\boldsymbol{r},\boldsymbol{r}',\omega).\boldsymbol{J}^{fl}(\boldsymbol{r}',\omega)d^3\boldsymbol{r}' \qquad (8)$$

The terms $\bar{\boldsymbol{G}}^E$ and $\bar{\boldsymbol{G}}^H$ are dyadic Green's functions for electric and magnetic fields, respectively. They relate the electric and magnetic fields of frequency, *ω*, observed at location $\boldsymbol{r}$ to a source at location $\boldsymbol{r}'$. The relationship between the electric and magnetic dyadic Green's functions is given by Eq. (9):

$$\bar{\boldsymbol{G}}^H = \nabla\times\bar{\boldsymbol{G}}^E \qquad (9)$$

Finally, the monochromatic radiative heat flux is calculated as the ensemble average of the Poynting vector, as given in Eq. (10):

$$\langle \boldsymbol{S}(\boldsymbol{r},\omega)\rangle = \frac{1}{2}Re\,[\langle \boldsymbol{E}(\boldsymbol{r},\omega)\times\boldsymbol{H}^*(\boldsymbol{r},\omega)\rangle]. \qquad (10)$$

The heat transfer coefficient between two media can be defined as shown in [33]:

$$h_r(d,T) = \frac{1}{4\pi^2}\int_0^\infty\int_0^\infty \frac{\partial\Theta}{\partial T}\xi_{12}(\omega,\beta)\beta d\beta d\omega \qquad (11)$$

where $\xi_{12}$ is the energy transmission coefficient between two media, $\beta$ is the parallel wavevector component, *d* is the separation distance and *T* is the temperature.

Furthermore, when the second term in equation (6) is dropped, the local density of the electromagnetic state can be expressed based on the vacuum energy density $\boldsymbol{U}_\omega\,(\boldsymbol{r},\omega)$ and $\boldsymbol{\Theta}(\omega,T)$ as:

$$\rho_\omega(\boldsymbol{r}) = \frac{\langle \boldsymbol{U}_\omega\,(\boldsymbol{r},\omega)\rangle}{\boldsymbol{\Theta}(\omega,T)} = \frac{\varepsilon_0\langle|\boldsymbol{E}(\boldsymbol{r},\omega)|^2\rangle+\mu_0\langle|\mathbf{H}(\boldsymbol{r},\omega)|^2\rangle}{\boldsymbol{\Theta}(\omega,T)} \qquad (12)$$

In addition, the net radiative heat flux can be calculated as:

$$q_{net} = \frac{1}{8\pi^3}\int_0^\infty[\boldsymbol{\Theta}(\omega,T_1) - \boldsymbol{\Theta}(\omega,T_2)]d\omega\int_0^{2\pi}\int_0^\infty \xi(\omega,\beta,\phi)\beta d\beta d\varphi \qquad (13)$$

where $\varphi$ is the azimuthal angle [33].

In the near-field radiative heat transfer regime, the contributions of surface modes and total internal reflection result in a radiative heat flux exceeding the blackbody predictions. The heat transfer in this regime can be enhanced by maximizing the photon transmission probability in the *ω–k* space. Here, *ω* refers to all accessible frequencies, and *k* is the parallel wave vector. The theoretical maximum heat transfer limit defined here is *1/d*, where *d* is the distance between two planar bodies. Different approaches may be applied to achieve considerable enhancement.

One approach is to choose materials that support surface waves. Surface waves, such as surface phonon polaritons and surface plasmon polaritons, can enable access to the large *k* channels, where *k* represents orders of magnitudes exceeding that of the wave vector, $k_0$, in free space. Another approach is to use nanostructures, such as photonic crystals, multilayer structures, metamaterials, meta-surfaces, and grooves, which allow the existence of multiple resonance modes. These designer structures result in the enhancement of radiative heat transfer at the near-field. In addition, widening the *ω–k* space through the use of hyperbolic dispersion can enhance the radiative heat transfer at the near-field. This improvement plays a vital role in nano-scale energy conversions, such as those in energy harvesting and radiative cooling applications [34–39]. Accordingly, in recent years, extensive research has been performed on the theoretical, numerical, and experimental analyses of radiative heat transfer in the near field in metamaterials and nano-photonic/nano-plasmonic systems [31,32,40–69].

Figure 1 displays how nanostructured materials alter the Planckian radiative transfer by showing a few exemplar concepts. Based on Wien's displacement law, we know that spectral radiance of black-body radiation per unit wavelength, peaks at the wavelength $\lambda_{peak} = b/T$, where $b$ = 2898 $\mu m.K$ and $T$ is the temperature of the surrounding medium measured in K. When the separation distance $d$ between media is less than or comparable to the $\lambda_{peak}$ , near-field radiative heat transfer is the dominant mode of heat transfer (in comparison to the far-field radiative heat transfer). Through the schematics of Fig. 1, we distinguish between far-field and near-field radiation by depicting the relation between $d$ and $\lambda_{peak}$. Figure 1(a) shows Planck's blackbody far-field radiative heat transfer regime. In this scenario, the media are separated from each other by a distance ($d$) greater than the thermal wavelength (λ). The medium with a higher temperature value acts as a heat source (emitter), and the other medium with a lower temperature value represents a heat sink (absorber). Figure 1(b) depicts the transition mechanism required to enter the near-field radiative heat transfer regime. In this regime, in Fig. 1(c), contributions from the surface and evanescent waves result in spectral enhancements of radiative transfer beyond the Planckian regime, as shown by the red curve in Fig. 1(e). Furthermore, extraordinary enhancements are demonstrated beyond the Planckian regime for the cases of topological radiative heat transfer, which is schematically represented in Fig. 1(d) and plotted in Fig. 1(e) in blue. For a clear comparison of the cases, Planck's distribution is also depicted in black in Fig. 1(e).

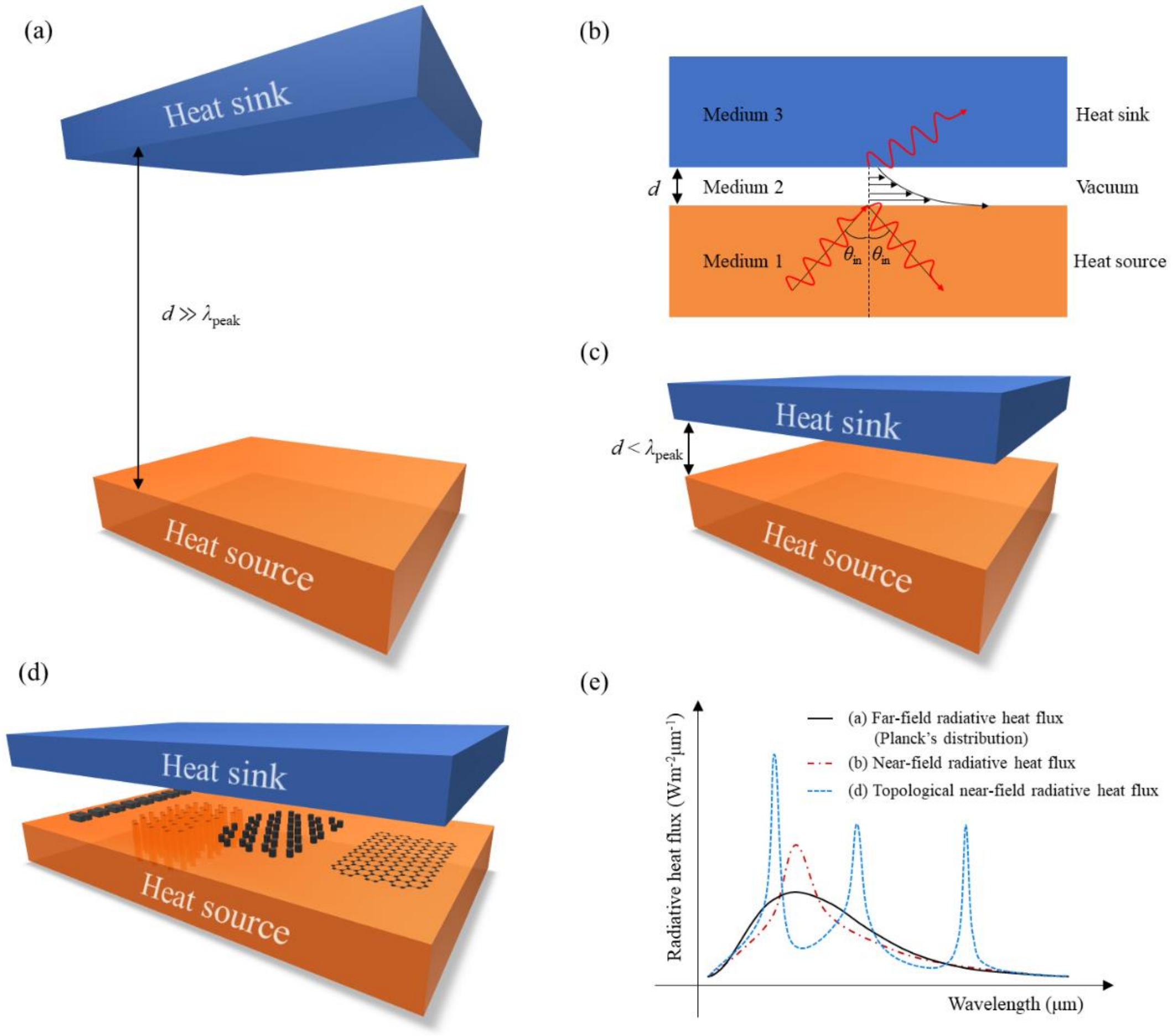

**Fig. 1**. Schematics of (a) Planck's far-field radiative heat transfer when the distance between the heat source and heat sink exceeds the thermal wavelength considered. (b) Transition mechanism from far-field to near-field radiative heat transfer depicting contributions of surface and evanescent waves. (c) Near-field radiative heat transfer. (d) Possible topological radiative heat transfer scenarios depending on 1D or 2D topological nano-photonic structures or materials of heat sink. (e) Possible radiative heat flux across various wavelengths for far-field, near-field, and topological radiative heat transfer scenarios.

Near-field thermal radiation refers to thermal electromagnetic radiation occurring at very short distances, where energy transfer rates are significantly higher due to phenomena such as evanescent waves and photon tunneling. The enhancement rate of near-field radiative heat transfer is heavily dependent on the material properties and geometries considered. Besides, commonly used systems include porous media, photonic crystals, metamaterials, and hyperbolic materials, the applications of the many-body theory in near-field thermal radiation have been widely studied [70–74]. Also, the fundamentals of radiative heat transfer along with recent experimental advances and novel theoretical approaches towards near-field radiative heat transfer in many body systems have been reviewed thoroughly in [75–77] available for interested readers.

In this context, the topological properties of materials, which remain unchanged under continuous deformation, are of crucial importance. Materials with these properties, such as topological insulators, possess edge states that are robust against certain types of disorder, resulting in enhanced and stable heat transfer. These materials can also tune the emission spectrum, enabling narrow-band thermal emission at specific wavelengths, which is advantageous for precise thermal management. Additionally, topological materials can exhibit non-reciprocal heat transfer, where the direction of heat flow is controlled, beneficial for the design of thermal diodes. The localized channels for heat transfer provided by topological surface states are ideal for applications requiring localized heating or cooling without affecting the bulk material. This relationship is central to advances in the thermal management of nanoelectronics, improving the efficiency of thermophotovoltaic devices, and enhancing thermal magnetic recording.

Recent advancements in topologically protected many-body systems reveal that thermal radiation can benefit from unidirectional heat transport, facilitated by edge-state currents in these systems. These unidirectional energy channels ensure the circulation of thermal energy in closed orbits, resulting in a non-zero angular momentum of the electromagnetic field [78]. Particularly, it has been demonstrated that heat flux circulates in directions led by incomplete electron cyclotron orbits, maintaining energy flow even at zero temperature [78].

In many-body topological systems, various unconventional thermal effects such as persistent heat fluxes [78,79], persistent spins, and angular momenta of both near-field and far-field thermal radiation [80,81], the Hall effect of thermal radiation on magneto-optical materials [82,83], and the dynamic control of magneto-optical materials via surface modes [84] have been explored. The topologically anomalous radiative Hall effect in Weyl semimetals [85] and radiative heat transfer in vibrational lattices, as well as between graphene nanoparticles, have been investigated within the framework of the Su–Schrieffer–Heeger (SSH) model [86]. These studies also encompass many-body quantum systems with SSH-type models or Dirac/Weyl semimetals and topological insulators with global symmetries such as time-reversal, particle-hole, and chiral symmetries.

We review the recent developments in the interplay of topological photonic and thermal radiation in the near field in many-body systems which emphasize the applications of periodic and aperiodic models, two-dimensional (2D) Dirac and magnetic Weyl semimetals, and topological insulator slabs with and without broken symmetries. These advances highlight the potential for new and efficient ways to manage thermal radiation, leveraging the unique properties of topological materials. We also highlight the possibility of achieving unidirectional and gigantic radiative heat transport at the nano-scale through the selective design of material, shape, size, and temperature properties for various potential applications, including thermal management assisted by an applied magnetic field. These topological effects are categorized into five groups: (I) SSH model, (II) Quasi-periodic nanoparticle chains, (III) Dirac and Weyl semimetals, (IV) Time-reversal or chiral symmetry breaking, and (V) Other 2D or 3D topological insulators. In each category, some of the fundamentals of the considered topological models are first reviewed; then, a review of existing supporting literature for each of them is provided to identify and discuss the interplay of topological photonics and super-Planckian regime of heat transfer in many-body systems. In conclusion, we offer our perspective on potential future directions for advancing the field of topological near-field radiative heat transfer.

## 2. SSH Model

Some of the fundamental ideas behind topological insulators can be understood using a simple one-dimensional (1D) model popularly known as the SSH model, which was originally developed to explain the electronic structures of poly-acetylene chains [87]. Among the wide variety of topological models, the SSH model typically describes the band topology in condensed matter physics and photonics [2,88]. Generally, this model consists of a 1D chain with N unit cells, each with two sublattices $A_m$ and $B_m$; spin-less particles, such as electrons or photons hop between these two sites with staggered hopping amplitudes or coupling coefficients, as depicted in Fig. 2(a). The Hamiltonian of this model is given by,

$$H_{SSH} = \sum_{m=-\infty}^{\infty} \left\{ \left( vA_m^{\dagger} + wA_{m+1}^{\dagger} \right) B_m \right\} + h.c. \quad (14)$$

where, *m* denotes the unit cell, *v* and *w* are the intra-coupling and inter-coupling coefficients, respectively, which are usually considered to be real and positive. The topological features of the SSH model depend on the staggered coupling coefficients $v$ and $w$; a bandgap gets created when $v \neq w$ and the bandgap is given by $2\,|w - v|$. Moreover, when $v > w$, the diatomic chain shows a bandgap at the band edge corresponding to a trivial topology with a 1D Berry phase *aka* Zak phase equal to 0. When $v < w$, a localized and topologically protected state in the bandgap emerges which is characterized by a non-trivial topological invariant with Zak phase equal to π. Changing the topological invariant in the SSH model, either requires closing and reopening the bulk gap or simply breaking the chiral symmetry present in the SSH chain. The nontrivial bulk topology ensures the presence of the edge states at the chain boundary. The topological edge states are robust against defects or other perturbations owing to the unchanged nontrivial topology of the bulk and the chiral symmetry that is preserved under perturbations.

### *2.1. SSH Model in Carbon-Based Nanomaterials*

In 2019, Tang *et al.* [86] reported an exclusive phenomenon of enhanced near-field heat transfer through the vacuum gap between carbon nanostructures with topological properties and explained the functions of electronic edge states by using standard SSH chains. Two types of carbon-based nanostructures are considered - the zigzag single-walled carbon nanotubes (SWCNT) and the graphene nanotriangles (GNT), respectively, which inherently support topological edge states. Authors have employed the non-equilibrium Green's function formalism within random phase approximation to estimate the heat current in such structures which is applicable within a range of 1.5 nm < $d \ll \lambda_{th}$. The corresponding variations of heat current $J$ (nW) with the vacuum gap separation $d$ (nm) as shown in Fig. 2(b) and (c) for SWCNT and GNT, respectively, clearly exhibit an enhanced heat current flow with the increasing gap separation. Interestingly, authors were able to explain such peculiar behavior of electronic edge states in carbon nanostructures like SWCNT and GNT contributing to the heat transfer in the light of the underlying physics associated with the heat transfer between simple SSH chains. The topological phase transition in the 1D SSH chain (as shown in Fig. 2(a)) occurs when the hopping parameter, $\lambda$, of electrons changes; when $\lambda > 0$, the SSH chain is in a trivial state, and a topologically non-trivial region appears when $\lambda < 0$. The energy spectrum of an open SSH chain with 160 lattice sites is depicted in the right inset of Figure 2(d) where the zero-energy gap states appear in the region with negative $\lambda$ values. The variation of the heat current, $J$ with the hopping parameter $\lambda$, for a vacuum gap separation, $d = 3$ nm, is shown in Fig. 2(d) which explicitly shows a large amount of heat current in the nontrivial region ($\lambda < 0$) compared to that in the topologically trivial ($\lambda > 0$) region. Moreover, a sharp transition is observed during the topological phase transition from trivial to nontrivial due to the existence of edge states in the non-trivial region that considerably contribute to the heat current.

Another interesting feature of such NP chains is the variation of the spectral transfer function, $\mathcal{T}(\omega)$ for a range of hopping parameters where $\lambda < 0$ (as shown in Fig. 2(e)) with a fixed $d = 3$ nm. Notably, authors have measured a critical gap distance $d_c$ for which the heat flux attains the maximum in the presence of edge state i.e., $\mathcal{T}(\omega \to 0) \approx 1$ around $d_c$. As the hopping parameter changes from $\lambda = -1$ to $\lambda = -0.22$, the heat current continues to increase as can be seen in Fig. 2(e). The 3 nm gap separation is the critical distance, $d_c$, corresponding to $\lambda = -0.22$, where the resonant peak of $\mathcal{T}(\omega)$ is at a frequency approaching $\omega = 0$. For $\lambda < -0.22$, the same gap separation is smaller than the corresponding critical distance, and the resonant peak appears at a higher frequency. When $\lambda$ exceeds –0.22, the peak of the spectral transfer function at a small angular frequency is observed to decrease as the trivial phase is approached; therefore, the heat current is also annihilated. By summarizing these results, one may conclude that the enhancement of the heat current in the presence of edge states can be utilized to design a near-field thermal switch provided tuning of the edge states is possible.

*2.2. 1D SSH chain of plasmonic nanoparticles*

The tremendous potential of SSH model-based photonic structures to couple and contribute to a considerable amount of heat flux enhancement for a shorter nanoparticle (NP) chain length has already been discussed in detail. Here, we discuss the exclusive behavior of the SSH-based NP chain for any arbitrary length of NPs that is longer than the thermal wavelength of interest. This surely is in contrast to the expectation based on the analogy of the super-Planckian near-field regime of radiative heat transfer and was reported recently by A. Ott *et. al.*[89] in which a topological SSH chain of 20 plasmonic indium antimonide (InSb) NPs of 100 nm radius and lattice constant of 1 μm are found to exhibit radiative heat flux through the chain in[89] the nontrivial topological phase. Figure 2(f) schematically represents the 1D SSH-based NP chain of InSb where $t$ is the spacing between two NPs A and B and is related to the lattice constant $a$ by the relation $t = \beta a/2$, where $\beta$ is the tuning parameter that determines the topological nature of the structure. The first NP is heated up to 310 K keeping the temperature of other NPs at 300 K. The variation of the spectral power received by the last NP with the angular frequency $\omega$ is depicted in Fig. 2(g) for both the longitudinal and transverse modes at two different $\beta$ values – 0.7 and 1.3 corresponding to the trivial ($\beta \leq 1$) and the nontrivial ($\beta > 1$) topological phases, respectively. It is explicitly shown that in the trivial topological phase ($\beta$ = 0.7), the heat flux is carried by the longitudinal and transverse band modes at lower and higher frequencies, respectively. In contrast, in the nontrivial phase ($\beta$ = 1.3) the dominant contribution to the spectral heat flux for both polarizations emanates from the edge modes. The degeneracy of the longitudinal and transverse edge modes is shown by the vertical dashed line in Fig. 2(g), in which the degenerate frequencies lie in the gap between the two bands. Notably, the dependency of the edge mode contribution to the heat flux on the length ($l$) of the NP chain is found to mimic the similar trend obtained for the contributions from band modes i.e., the heat flux varies as $1/l^2$ and $1/l^4$ for transverse and longitudinal modes, respectively, provided that the number of NPs exceeds 8. Thus, the topological edge modes, although confined at the edges, unexpectedly provide a major radiative heat flux channel along with the band mode-induced channel and explain the domination of the edge-mode transported heat flux for any given arbitrary NP chain length. Such striking behavior is likely to occur owing to the presence of retardation in the SSH model which results in the long-range coupling between the first and last NP in the given chain. Such observation of the edge mode-dominated heat flux reveals that topological edge modes can play an important role in heat flux channels along the SSH chain through its topological phase, laying the foundation for the future exploration of near-field radiative heat transfer in topological systems.

*2.3. 2D SSH lattice of plasmonic nanoparticles*

Building upon the discussions of the previous section, A. Ott *et al.* [90] conducted an in-depth investigation of the enhanced near-field energy density of the edge states in topological many-body systems during the topological phase transition. The study essentially reports a generalized expression for both the electric and the magnetic near-field energy density of $N$ dipoles each with specific temperatures while the background temperature is maintained at a fixed value other than the dipoles. The authors used the formalism to study the spectral energy density of edge states in the 1D SSH model-based InSb NP chain as well as of the corner states in 2D SSH lattice-based plasmonic InSb NPs. Figure 2(h) and (i) depict the spectral energy density in the vicinity of 1D SSH InSb NP chain for $\beta > 1$ and $\beta < 1$, respectively, in which a clear enhancement of energy density at the edges of the chain is seen in Fig 2(h), whereas a uniform energy distribution is observed in Fig. 2(i). Similarly, Figure 2 (j) and (k) exhibit the variation of energy density in a 2D SSH lattice with 36 NPs for $\beta > 1$ and $\beta < 1$, respectively. The domination of corner states is explicitly observed in Fig. 2(j) whereas uniform energy density is shown in Fig. 2(k). These results are clear proof that topological phase transition eventually enhances the energy in the vicinity of the edge and corner of the structures, which also makes them experimentally accessible with state-of-the-art near-field thermal profilers such as thermal infrared near-field spectroscopy, thermal radiation scanning tunneling microscopy, and scanning noise microscopy for further investigation. Moreover, authors have also reported the existence of defect-immune edge and corner states in such SSH model-based photonic structures by using their proposed formalism as long as the defects are away from the edge and the corners concerned.

In this context, the same group of researchers has extended their investigation on the impact of topological edge states on near-field radiative heat flux in a honeycomb lattice of InSb NPs with both the $C_6$ and inversion symmetry [91]. They have theoretically shown that the edge and corner NPs are heated through the channel provided by the edge and corner topological states, whereas the heating of bulk NPs results from bulk mode propagation. It is also shown that a reduced damping constant results in the domination of edge modes over bulk modes for enhanced heat flow.

*2.4. SSH model for terahertz temperature sensing*

Lately, such SSH model-based nanostructures have been investigated further for thermal sensing applications in the terahertz (THz) regime. To this end, in 2020, Wang *et. al.* [92] reported a detailed study on the topological plasmon polaritons (TPPs) and their substantial effect on radiation generated at the THz level. The authors have theoretically demonstrated the photonic band structure and eigenmode distribution of a 1D dimerized InSb microsphere chain, which exhibits properties analogous to the topological edge state in the 1D SSH model. The schematic of the system considered in their work is depicted in Fig. 3(a). The system is composed of microspheres of InSb with a radius of 1 μm, whereas $d_1$ and $d_2$ are the two unequal spacings of the two different lattice sites such that $d = d_1 + d_2 = 10$ μm, and the dimerization parameter is $\beta = d_1/d$. For a complete understanding of the system's behavior, the authors have employed a full picture of the near-field and far-field dipole-dipole interactions beyond the traditional nearest-neighbor approximation in the conventional SSH model to study the band structure and eigenstate distributions of the longitudinal modes. In this context, the transverse modes are not being considered due to their weaker localization rate, rendering them difficult to detect experimentally and these findings are thoroughly reported in their previous works [93,94]. Notably, the estimated eigenfrequencies of such a system are found to be complex owing to the non-Hermiticity of the system, and also the Zak phases are complex-valued and quantized (0 or π) irrespective of the broken chiral symmetry of the system. Moreover, it perfectly obeys the conventional bulk-boundary correspondence and other topological properties as in the case of standard 1D dimer chains. There are several intriguing features of such TPPs, like (a) they exhibit quite a significant variation of resonant frequencies $\omega_{TPP}$ (0.2 to 4 THz) over a range of temperature $T$ change (160 to 350 K), (b) stability of the topological modes against the geometric parameter changes, and (c) the bandgap enhancement with increasing temperature, which make them very much suitable for experimental temperature sensing applications. Authors have rigorously shown that the variation of the temperature susceptibility parameter $S = \Delta\omega_{TPP} \,/\, \Delta T$, where Δ indicates a small variation in a physical quantity, goes higher with the increase in temperature. To further establish their claim, a figure of merit is defined as $FoM = ST/\Gamma = \Delta\omega_{TPP}T/(\Gamma\Delta\Gamma)$ which also exhibit a value of 150 at room temperature which is much higher than that obtained for any other state-of-the-art systems. The corresponding variations of susceptibility $S$ and $FoM$ with temperature T quantifying the temperature sensing performance of TPPs are shown in Fig. 3(b). Further, the experimental detection of the topological photonic modes can be performed in the THz domain by using scattering-type scanning electron microscopy (s-SNOM) which very closely follows the theoretical estimations of LDOS distribution for the system. In Fig. 3, a detailed analysis of the LDOS distribution of both the topological and non-topological chains is shown with a slight temperature variation which explicitly shows that for topological chains ($\beta = 0.7$) the resonant peaks are at the center of the bandgap whereas for non-topological chains ($\beta = 0.3$) they are slightly shifted towards the band edge. Moreover, a temperature change of 1 K results in a significant amount of resonant peak shift. Notably, the robustness of the topological mode frequencies against any geometric disorder is incomparable as can be seen from Fig. 3(d), where the same amount of disorder severely distorts the non-topological mode frequencies but it has no effect on the topological one. In addition, authors have also provided a proper calibration method by using the ratio of LDOS of the topological chain to that of the non-topological chain for an accurate estimation of LDOS spectrum for the topological mode frequencies and thus temperature susceptibility which certainly considers the non-zero LDOS of the non-topological modes arising from the bulk modes. Such findings are surely exclusive and eventually open up a new avenue for terahertz temperature sensing employing the topological photonic devices that provide a robust temperature sensitivity owing to the defect immune topological TPP modes.

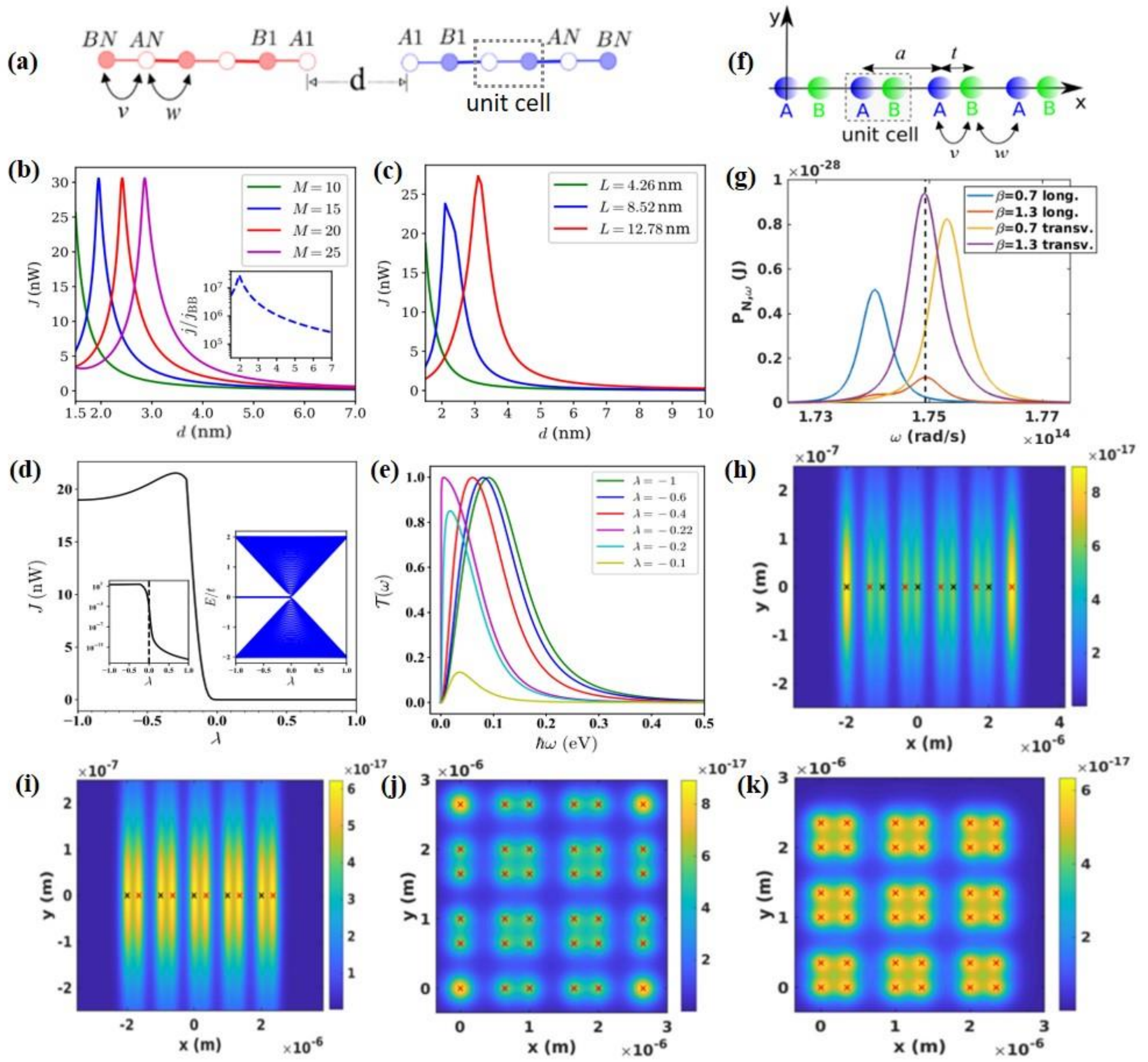

**Fig. 2**. (a) Schematic representation of 1D SSH model. Variation of heat current with the gap separation *d* for (b) SWCNT, and (c) GNT. (d) Heat current versus hopping parameter, $\lambda$, for gap separation $d$ = 3 nm. The inset shows the spectrum of the SSH chain with 160 lattice sites as a function of $\lambda$ [86]. (e) Variation of the spectral transfer function at different hopping parameters $\lambda$ with a fixed $d$ = 3 nm. (f) Schematic representation of the 1D InSb NP chain with lattice constant *a* and unit cell dimension *t*. (g) Spectral power for longitudinal and transversal modes with $\beta$ = 0.7 and $\beta$ = 1.3 with the temperature of the first NP set at 310 K and that of all other particles at 300 K [89]. Spectral energy density for 1D SSH chain of 10 nano-particles for h) non-trivial case where $\beta$ = 1.3 and i) trivial case where $\beta$ = 0.7 [90]. Spectral density of 2D SSH lattice chain of 36 nano-particles for j) non-trivial case where $\beta$ = 1.3 and k) trivial case where $\beta$ = 0.7 assuming that all the NPs are at a temperature 350 K with the background set at 300 K [90]. Reproduced with permission: ©2019 American Physical Society (a - e), ©2020, American Physical Society (f, g), and ©2021 American Physical Society (h - k).

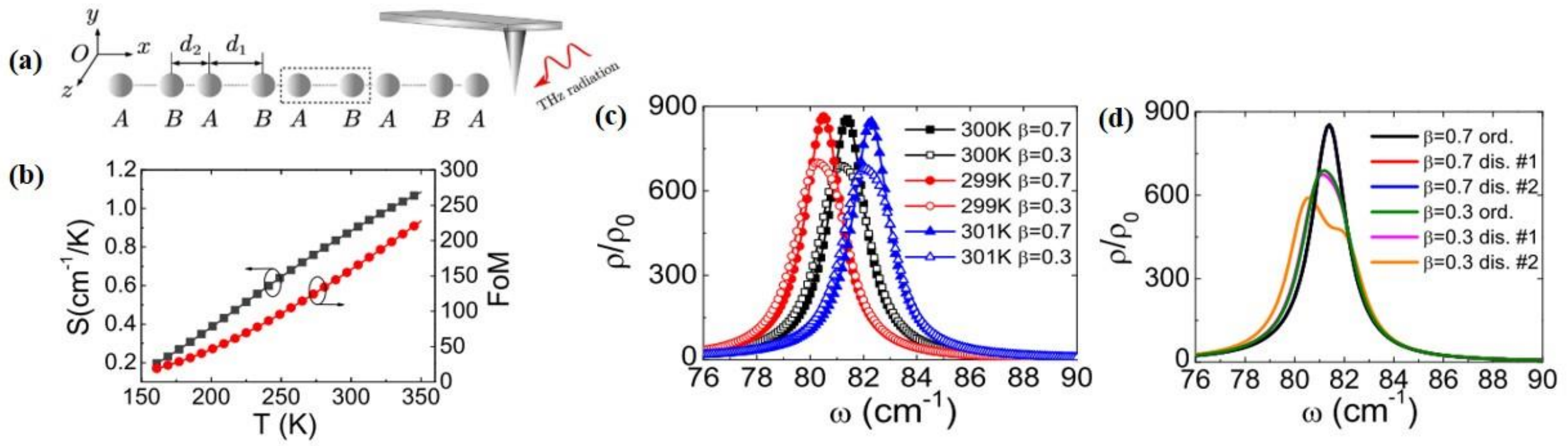

**Fig. 3.** (a) Schematics of InSb microsphere chain in the presence of scanning near-field optical microscope and terahertz radiation. (b) Variation of temperature susceptibility *S* and figure of merit with temperature *T* of the topological chain. (c) LDOS for topological and non-topological chains at temperatures 299 K, 300 K, and 301 K [92]. (d) LDOS at *T* = 300 K for topological and non-topological disordered chains versus ordered chains [92]. Reproduced with permission: ©2020 American Physical Society (a – d).

## 3. Quasi-periodic nanoparticle chains

Besides the evidence of topological features in various periodic lattice arrangements such as SSH model-based NPs, a quasi-periodic 1D geometry offers unique topological properties that eventually compel similar intense light localization by topological protection, essentially replacing the necessity of 'true' disorder in 1D lattices. As illustrated in literature, a quasi-periodic crystal is ordered but not periodic and also contains unusual rotational symmetries (like five-fold), unlike conventional crystals. Moreover, such quasi-periodicity maintains a ratio between two adjacent lattice distances called a golden ratio (like the one in the Fibonacci series), which is typical of its lattice structure. Based on this fundamental concept, a particular model namely the Aubry-André-Harper (AAH) model has been developed that exhibits an exact quasi-periodic 1D lattice geometry via a cosine modulation of their hopping parameter resulting in an incommensurate and commensurate model, and is expressed as,

$$x_{n+1} - x_n = d[1 + \eta\xi_n] \tag{15}$$

where, $\xi_n = \cos(2\pi\beta n + \varphi)$ , is the cosine modulation with $\beta$ being the periodicity of the modulation, and $\varphi$ is the modulation phase that corresponds to the momentum (wavenumber) in a synthetic orthogonal dimension extending the model to 2D; also, $x_n$ is the position of the $n^{th}$ lattice site, $d$ is the periodic lattice constant without modulation, and $\eta$ controls the amplitude of distance modulation. AAH model exhibits a metal-to-insulator phase transition based on a critical modulation point. The modulation of the structural periodicity imposes a lack of inversion centre in the model that eventually results in nontrivial topology. Moreover, a metallic phase of the model has all the eigenstates delocalized, whereas the insulator phase supports localized eigenstates. An incommensurate AAH model supports a topological phase transition from an extended light state to a localized one due to the onsite cosine modulation crossing the critical point responsible for such transitions [95], whereas a commensurate model has equal potential to explore novel topological states of light in such geometries [96]. Unlike SSH topology, a disorder in some AAH topologies may lead to a phase transition from topologically trivial to a topologically nontrivial phase [97]. Implementation of such quasi-periodicity in photonic structures, though, may lead to futuristic photonic devices for various targeted applications, and researchers around the globe have already investigated this domain rigorously.

In 2023, the research group of B. X. Wang and C. Y. Zhao proposed such a quasi-periodic model-based 1D array of Silicon Carbide (SiC) NPs supporting topological phonon polaritons (TPhP) for near-field radiative heat transfer [98,99]. In their first work [98], the 1D SiC NP array is given a quasi-periodicity along the x-axis mimicking the AAH model as given in Eq. (15). Figures 4(a) and (b) depict the distribution of nanoparticles described by Eq. (15) with three different values of modulation phase $\varphi$ for the incommensurate and the commensurate cases, respectively, keeping $d$ = 0.6 µm, and $\eta$ = 0.3. The authors have rigorously shown the bandgap structures, the existence of TPhPs within the bandgaps, and spectral heat transfer through the TPhPs for a particular modulation phase $\varphi$ and modulation period $\beta$ . Figure 4(c) depicts the longitudinal band structure for an incommensurate ($\beta = (\sqrt{5}-1)/2$) SiC lattice with 100 NPs as a function of the modulation phase $\varphi$ where the topological states within the bandgap are clearly seen. The net spectral heat transfer from the first NP to the last NP for three different $\varphi$ is shown in Fig. 4(d) while the other parameters are $d$ = 0.6 µm, and $\eta$ = 0.3. It has been assumed that the first NP is at 310K while other NPs are at 300K. For $\varphi$ = 0.2π, no TPhPs are present within the resonant bandgap as explicitly shown in Fig. 4(c), and the heat transfer occurs through bulk modes only, whereas for $\varphi$ = π, and $\varphi$ = 1.2π, TPhPs are supported by the respective bandgaps which enhance the net heat transfer compared to that through the bulk modes. Also, the net heat transfer increases if more topological edge states are present closer to the resonant frequency of phonon polaritons for SiC. Notably, for an even number of lattice sites (100 NPs), the edge states are localized at both the boundaries of the array, whereas for an odd number of lattice sites (99 NPs), they are localized at any one boundary of the array. However, this does not affect the TPhP-dominated heat transfer. Further, the effect of commensurate lattice (rational $\beta$) on heat transfer has also been presented as shown in Figs. 4(e) to (h) with two different $\beta$ values. For $\beta$ = ½, the band structure is shown in Fig. 4(e). The net TPhP-aided heat transfer has been observed for only $\varphi$ = π (Fig. 4(f)), as there is no proper bandgap for this commensurate lattice except a set of zero-energy modes near the resonance frequency that excites the TPhPs as can be clearly seen from Fig. 4(e); for $\varphi$ = 0.2π, the heat transfer is bulk mode dominated. In Fig. 4(g), the band structure for a 100 NP chain with $\beta$ = ¼ is shown. Figure 4(h) explicitly depicts a similar trend of heat transfer at three $\varphi$ values (corresponding to the states in Fig. 4(g)) that the presence of TPhPs in the bandgaps enhances the radiative heat transfer compared to that through bulk modes. They have also shown that the lower the damping coefficient, the higher the heat transfer rate in such a TPhP-aided quasi-periodic NP array.

In the same year, authors reported near-field radiative heat transfer in another quasi-periodic lattice structure, namely interpolating the Aubry-André-Fibonacci (IAAF) model which combines both the Aubry-André (AA) model and the Fibonacci chain, thereby extending the limit of quasi-periodic 1D geometries [99]. For this, a modified form of Eq. (15) has been used where $\xi_n$ takes a more general form, $\xi_n = \frac{\tanh\{\tau[\cos(2\pi\beta n+\varphi)-\cos(\pi\beta)]\}}{\tanh\tau}$ , which describes the distance modulation of the NP array combining AA and Fibonacci models. The radiative heat transfer has been studied in a 100 NPs SiC chain at three limiting cases with $\tau \rightarrow 0$ corresponding to the AAH limit, $\tau \rightarrow \infty$ corresponding to the Fibonacci limit, and an intermediate case of $\tau$ = 5 for the IAAF model. For $\tau$ = 0.01 i.e., at the AAH limit, a similar net heat transfer rate, as shown in Fig. 4(d), for an incommensurate AAH lattice has been observed. With $\tau = 10^6$ i.e., at the Fibonacci limit, the band structure, as shown in Fig. 4(i), exhibits flat mid-gap zero energy states with a sharp transition from the center of the gap to the edge, unlike the AAH model. Figure 4(j) depicts the spectral heat transfer rate of 100 NP chains at $\varphi$ = 0.1$\pi$ and $\varphi$ = 1.1$\pi$ in which the latter exhibit TPhP-dominated maximum heat transfer. Interestingly, the effect of an odd number of lattice sites (99 NP chain) on net heat transfer in the Fibonacci limit is more pronounced as can be observed from Fig. 4(k), which also appears to follow the trend of sharp transition of the topological states from center to edge of the bandgap. Such topological states in odd-numbered chains are also found to be closer to the resonant frequency of SiC NP.

Further, the IAAF model-based 100 NP chain exhibits an intermediate characteristic between the AAH chain and Fibonacci chain with a smoother transition within the gap except for sharp changes near the boundaries. Also, the net heat transfer rate resembles that of the two previous limiting cases where the presence of TPhPs enhances the heat flow from the first to the last particle. Notably, the effect of an odd number of lattice sites on enhanced heat transfer with mid-gap topological states close to the resonance frequency compared to that in the even-numbered NP chain is shown in Fig. 4(l). Authors have envisaged that such quasiperiodic structures contain rich physics that may have significant potential for devising thermal sensors, routers, photovoltaics, etc. Moreover, such topological states may exhibit strong coupling with the environment comprising polar material substrates or metasurfaces.

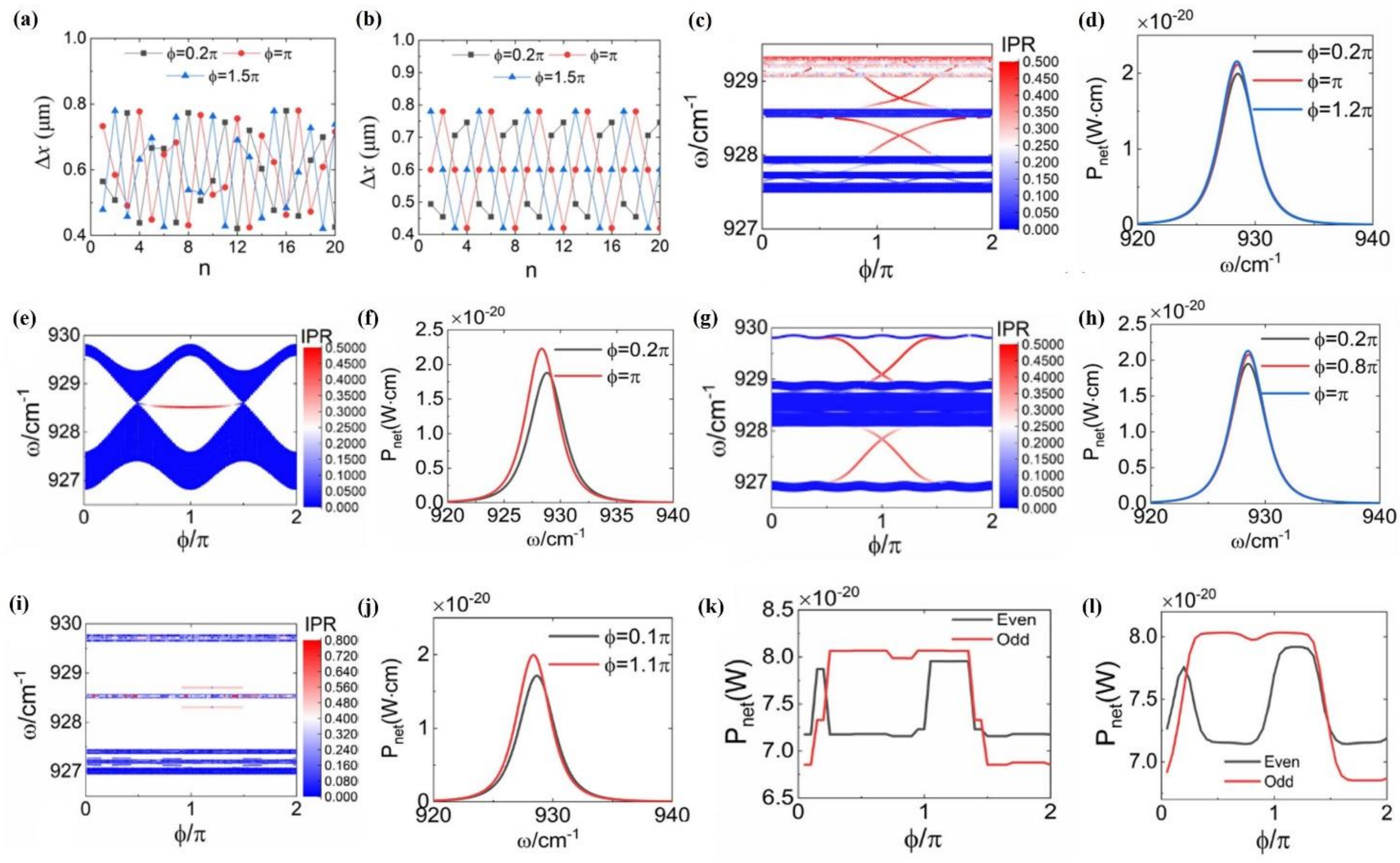

**Fig. 4**. The distribution of NPs with different modulation phase $\varphi$ for (a) incommensurate case $\beta = (\sqrt{5}-1)/2$, and (b) commensurate case $\beta = 1/4$. The other parameters are d = 0.6 μm, and $\eta$ = 0.3[98]. The longitudinal band structure as a function of $\varphi$ for (c) an incommensurate AAH lattice [98], a commensurate AAH lattice [98] with (e) $\beta$ = ½ and (g) $\beta$ = ¼, (i) a Fibonacci chain,[99] with $d$ =0.6 μm, $\eta$ = 0.3, 100 SiC NP. The spectral heat transfer rate at different $\varphi$ corresponding to bulk modes and topological phonon polaritons for (d) an incommensurate AAH lattice [98], a commensurate AAH lattice [98] with (f) $\beta$ = ½ and (h) $\beta$ = ¼, (j) a Fibonacci chain,[99]. The comparison of net heat transfer as a function of $\varphi$ between even-numbered chain (100) and odd-numbered chain (99) for (k) Fibonacci model [99], and (l) IAAF model [99]. In all the cases, it is assumed that the temperature of the first NP is set at 310 K and that of all other particles at 300 K. Reproduced with permission: ©2023, American Physical Society (a-h); ©2023, Elsevier (i-l).

## 4. Dirac & Weyl semimetals

Dirac and Weyl semimetals lie in the category of 2D/3D gapless topological phase where the degenerate point between two bands is held inside the bulk instead of at the interface like topological insulators. Dirac materials have a cone-type low-energy band structure (Dirac-like) within the first Brillouin zone that allows linear energy dependence on momentum. Graphene is the most well-known material with such Dirac cone bands at the K and K' points. In contrast, Weyl semimetals have an electronic band structure with single degenerate bands that have bulk band crossings (called Weyl points) [18]. Around the Γ momentum points, the Weyl Hamiltonian has a 4 × 4 matrix, generating chiral or Weyl fermions with a band structure. Weyl semimetals exhibit robust topological nodes on highly conducted surfaces [100]. To characterize the topological properties of such materials, Chern number is often used for Dirac or Weyl materials. When the material has insulating properties in bulk and edge currents that do not require an external magnetic field, the material is called a Chern insulator. When two identical planar materials are closely separated from each other, the interaction energy between the two at thermal equilibrium exhibits typical scaling laws, such as $1/L^2$ for three-dimensional (3D) metals and insulators, $1/L^4$ for 2D insulators, and $1/L^{5/2}$ for 2D metals, where $L$ is the distance between the two materials. The work of Rodriguez-López et al [101] on the framework of local approximations investigated the short-distance asymptote of the near-field heat transfer between two identical Dirac materials, such as topological Chern insulators and graphene. They demonstrated that two sheets of Dirac materials with a separation distance, $L$, show a $1/L$ scaling law within the near-field regime. In Fig. 5(a), the density plot of the transmission coefficient in the evanescent regime is depicted for two identical 2D Chern insulators at $T_1 =$ 10 K and $T_2 = 1$ K; the plot only shows the evanescent regime where $K > \omega/c$. In this regime, the contribution of the propagating modes to the transmission coefficient is negligible. The transmission coefficient in the near-field regime is observed to vanish for frequencies that are less than $2\Delta/\hbar$ (where $\Delta$ is the Dirac insulator band gap, and $\hbar$ is the reduced Planck's constant), which exceeds the thermal envelope (shown by the horizontal dashed line in Fig. 5(a)). In the past years, extensive research on graphene has shown the possibility of tuning heat transport via chemical doping or electromagnetic gating, benefiting from the existence of surface plasmon polaritons from the terahertz region to the near-infrared region. The authors also studied the near-field asymptotes of radiative heat transfer in graphene. They derived the near-field asymptotes of the heat transport in graphene and determined how this transport scales with distance $L$. Figure 5(b) shows the transmission coefficient, $T(\omega, K)$, for two identical graphene sheets separated by $L$ = 10 nm. Furthermore, the radiative heat transfer between two identical 2D Chern insulators as a function of their separation distance was investigated. Their study results are shown in Fig. 5(c). The black curve represents the numerical evaluation of the evanescent contribution only. The dotted red line depicts the far-field Stefan–Boltzmann law and the green line shows the analytical expression yielding the $1/L$ asymptote.

In addition, Figure 5(d) depicts the results of the derived radiative heat transfer between two identical graphene sheets as a function of their separation. One sheet was at $T_1 = 400$ K, and the other was at $T_2 = 300$ K. The black line in the figure is the numerical separation, the dotted red line is the far-field Stefan–Boltzmann law, and the blue curve is the evanescent contribution with a scaling of $1/L^3$. The green curve corresponds to the analytical expression. They also discussed the validity range of their derivations when spatial dispersion was included or neglected. They noted that their derivation of the $1/L$ scaling law no longer applies if spatial dispersion was also considered. The non-local effects start appearing once the separation distance reduces to less than 0.1 nm. These findings on the optical behavior of Dirac materials enable reaching the regime where spatial dispersion effects, such as Coulomb drag, are crucial in analyzing the quantum interactions of materials.

### *4.1. Thermal Radiation in Magnetic Weyl Semimetals*

In the context of Weyl topological models for thermal radiation, the following work proposes a novel approach to achieve non-reciprocity. Without requiring the application of an external magnetic field, the approach is virtually a complete violation of Kirchhoff's law. Zhao et al. [102] studied the axion electrodynamics in magnetic Weyl semimetals and showed that it can be an effective method for creating non-reciprocal thermal emitters. The axion term emanates from the direct coupling between the electric and magnetic fields, that is, $\vec{E} \cdot \vec{B}$ in the Chern–Simons effective field theory [103]. The violation of Kirchhoff's law is known to be only possible by applying a strong magnetic field (approximately 3T) through magneto–optical effects or using narrow-band resonances utilized under a magnetic field (approximately 0.3T). However, the authors demonstrated that even without applying a magnetic field, breaking Kirchhoff's law is possible. Moreover, a non-reciprocal thermal emitter that is functional over a broad frequency range without an external magnetic field can be constructed because of the axion term in the Hamiltonian of the corresponding magnetic Weyl semimetals. The band structure of the proposed magnetic Weyl semimetal with nodes of opposite chirality at a distance of $2b$ in 5(f), the authors designed a non-reciprocal thermal emitter that can virtually completely violate Kirchhoff's law at a wavelength of approximately 15 μm. K-space is shown in Fig. 5(e). As illustrated in Fig. 5(f), A periodic array of gratings along the *x*-axis exists in the considered structure. The periodicity is 7 μm, and the strip width is 3.5 μm. The gratings, each 0.5-μm thick, were placed on a thick substrate. Owing to the chemical potential properties of the material, the magnetic Weyl semimetal is highly temperature-dependent. The structure was considered opaque, and the transmission was zero. Here, $b$ is along the $z$ direction, and the incident plane is the $xy$ plane; the angle of incidence is $\theta$. The band dispersion and continuum region of bulk modes (light blue) and the light cone of vacuum (light grey) are depicted in Fig. 5(g). The asymmetric band structures of surface plasmon polaritons outside the light cone and continuum region of bulk modes can be observed in this figure. Furthermore, the real component of the permittivity tensor for $T$ = 300 K and $T$ = 400 K was investigated (results not presented here). At T = 400 K, the difference between absorptivity and emissivity is remarkably reduced, indicating that the non-reciprocal effect in the designed thermal emitter is highly temperature-dependent. Such a strong temperature dependence of non-reciprocity may be used to switch the non-reciprocal thermal emission on or off for a selective frequency application or to build thermal rectifiers [105]. The authors argued that the violation of Kirchhoff's law persists over wide angular and frequency ranges, and the non-reciprocal radiative properties are highly temperature-sensitive. Their work suggests that topological quantum materials can be used as a solution to the long-standing challenge regarding materials in the construction of non-reciprocal thermal emitters. Accordingly, considerable opportunities in energy harvesting and heat transfer applications are introduced.

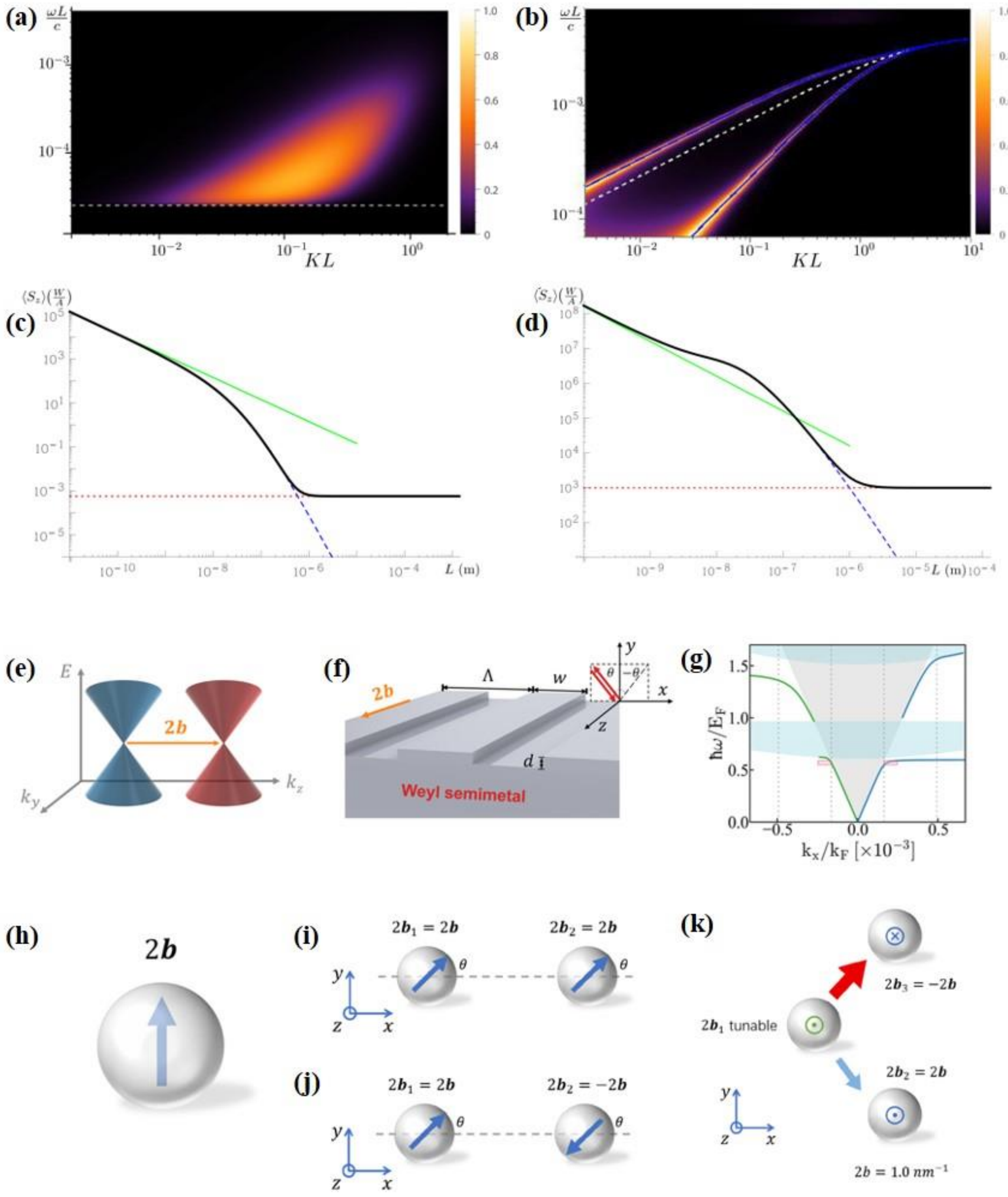

**Fig. 5**. (a) Transmission coefficient for two identical 2D Chern insulators in the near-field region with $T_1 = 10$ K and $T_2 = 1$ K. (b) Transmission coefficient for two identical graphene sheets. (c) Radiative heat transfer between two identical 2D Chern insulators as a function of their separation distance. (d) Radiative heat transfer between two graphene sheets as a function of their separation distance with $T_1 = 400$ K, and $T_2 = 300$ K. [84]. (e) Band structure of a magnetic Weyl semimetal. (f) Proposed photonic crystal for nonreciprocal thermal emission. (g) Band dispersion with continuum region of bulk modes (light blue) and light cone of vacuum (light grey); asymmetric band structures of the surface observed [102]. (h) Thermal radiation from a magnetic Weyl semimetal-based nanosphere. Radiative heat transfer between two identical Weyl semimetal-based nanospheres for (i) Parallel and (j) anti-parallel configurations. (k) Radiative thermal router consisting of three spheres made of magnetic Weyl semimetals[104]. Reproduced with permission: ©2015, IOP Publishing Ltd. (a)–(d); ©2020 American Chemical Society (e)–(k).

### *4.2. Weyl semimetal-based near-field thermal router*

The research on Weyl semimetal-based radiative transfer is further extended theoretically in the next work. A thermal router based on tuneable magnetic Weyl semimetals with potential applications for energy harvesting and heat transfer is proposed by Guo *et. al*.[104] Authors have both theoretically and numerically demonstrated a radiative thermal router based on tuneable magnetic Weyl semimetals. They used the excellent tunability and inhomogeneity of optical gyrotropy in magnetic Weyl semimetals to devise a selective thermal router. The proposed thermal router system consists of three magnetic Weyl semimetal spheres. Of the three spheres, one is tuneable, and the other two are fixed. The authors further explained the tunability and fixed nature of the Weyl semimetals as follows: the optical gyrotropy of Weyl semimetals is estimated by their band topology, particularly their position in the Brillouin zone. For this reason, their gyrotropic response is tuneable by moving the Weyl points around the Brillouin zone. In most Weyl semimetals, the position of Weyl points in the Brillouin zone is fixed; hence, they are referred to as fixed Weyl semimetals. In contrast, Weyl semimetals also have Weyl nodes that can be moved easily under a moderate external field; these semimetals are tuneable. The authors proposed a system that could selectively direct heat flow to a drain by moving the nodes in a tuneable magnetic Weyl semimetal using an external electric, magnetic, or optical field. Here, the band structure of a magnetic Weyl semimetal with two Weyl nodes of opposite chirality separated by $2b$ is identical to that shown in Fig. 5(i). The authors first considered thermal emission from one nano-sphere (N = 1) made of a magnetic Weyl semimetal with a tuneable Weyl node separation of $2b$ (Fig. 5(h)). It was observed that the resonant frequencies experience an increased splitting as $2b$ increases. Notably, among the three modes, one is strongly $b$-dependent, and the other two are $b$-independent. The $b$-dependent modes are associated with the gyrotropic response, and the $b$-independent mode depicts zero angular momentum. Figures 5(i) and 5(j) depict two different scenarios for radiative heat transfer between two nano-spheres made of magnetic Weyl semimetals. Figure 5(i) shows the two magnetic Weyl semimetals separated from each other at a distance of 320 nm (center to center) with a node separation of $2b_1 = 2b_2$, representing a parallel configuration. In Fig. 5(j), the separation distance is the same, but $2b_1 = -2b_2$, showing an anti-parallel configuration.

It is also observed that at an angle $\theta$ = 90°, the thermal conductance, $G_{12}$, for the anti-parallel configuration is extremely large compared with $G_{12}$ for parallel configuration owing to the efficient photon tunneling through the frequency-matched modes with opposite rotational direction as for the antiparallel case. The fields of the modes in the two particles move in the same direction; hence, they are approximately phase-matched. Notably, it tends to depend on the variation of $2b_1$, indicating that the structure could achieve the required functionality of radiative thermal routing. In addition, the authors verified that the thermal router remained robust to moderate variations in the radii of the nano-spheres or side lengths of the isosceles triangle. These findings could find applications in topologically protected thermal transport at the nano-scale level.

### *4.3. Anomalous Thermal Hall effect in Weyl Semimetals*

Here, we review the very recent work of A. Ott *et. al.*[85] demonstrate the many-body interactions induced by thermal photons in Weyl semimetals and that they could lead to an intrinsically originated anomalous Hall flux. Despite the photon thermal Hall effect, where the application of an external magnetic field is necessary for the anomalous thermal Hall effect, Weyl semimetals do not require the application of an external magnetic field. Notably, in the infrared region, Weyl semimetals exhibit an intense broadband non-reciprocal response owing to its non-reciprocal permittivity tensor. This response has a relatively large magnitude for the thermal Hall flux that are associated with their topologically non-trivial states leading to persistent heat current, angular momentum, and spin. Figure 6(a) schematically represents the Weyl semimetal network with four identical nanospheres positioned at four terminals forming a $C_4$ symmetry with $d = 300\,nm$, and the nanosphere radius $r = 100\,nm$. The three dipolar resonances are at $\omega_m = +1 = 1.486 \times 10^{13}$ , $\omega_m = 0 = 7.714 \times 10^{13}$, and $\omega_m = -1 = 3.865 \times 10^{14}\,rad/s$ . The non-reciprocal behaviour of the Weyl semimetal particles is presented by analysis of the transmission coefficient between particles *i* and *j* which is given as [85],

$$\mathcal{T}_{ij}(\omega) = \frac{4}{3} K_0^4 ImTr\left[\alpha_j\, G_{ji}\, \frac{\alpha_i - \alpha_i^{\dagger}}{2i}\, G_{ji}^{\dagger}\right], \tag{16}$$

where $k_0 = \omega/c$ , and $G_{ji}$ represents the dyadic Green's tensor between the $i^{th}$ and $j^{th}$ particles in the N-dipole network system; $\alpha_i$ represents the polarizability tensor of the $i^{th}$ particle. As a consequence of nonreciprocity, $\mathcal{T}_{ij} \neq \mathcal{T}_{ji}$ is expected which is an important concept for the existence of an anomalous photon thermal Hall effect. In Fig. 6(b), the asymmetry of the transmission coefficient is shown for the Weyl semimetal network. The asymmetry of clockwise and anti-clockwise heat fluxes can be observed in the inset by the ratio indicating the asymmetry, $\mathcal{T}_{31} \neq \mathcal{T}_{13}$. This asymmetry is due to non-reciprocity, which leads to a circular heat flux and is crucial to the thermal Hall effect. Authors have presented the magnitude of the Hall effect in terms of the thermal conductance $G_{ij}$, using the configuration of $C_4$, expressed as, $R_H = \frac{G_{13} - G_{31}}{G_{13} - G_{31} + 2G_{34}}$, where, the thermal conductance is given by [85],

$$G_{ij} = 3 \int_0^{\infty} \frac{d\omega}{2\pi} \frac{\partial \Theta(\omega, T)}{\partial T}\bigg|_{T = T_j} \mathcal{T}_{ij}(\omega). \tag{17}$$

In the expression of $R_H$, the numerator $G_{13} - G_{31}$ exhibits the asymmetry in the clockwise and anti-clockwise heat flows which is introduced by the non-reciprocal Weyl semimetal and is the origin of the anomalous photon thermal Hall effect. Interestingly, at higher temperatures, the anomalous photon thermal Hall effect becomes stronger; however, the strength of this effect can be decreased in the far-field regime and change its direction.

### *4.4. Dirac and Weyl semimetals for active heat flux modulations*

In 2021, G. Xu *et. al.*[106] demonstrated that the presence of a Dirac semimetal (DSM) in a simple emitter-receiver structure can highly influence the near-field radiative heat flux which in turn helps to actively modulate the heat flux through tuning of geometric parameters, Fermi level as well as the degenerate factor of 3D Dirac points. Active modulation of the radiative heat flux is important for heat transfer management like fast enhancement or suppression of heat flux by some parameter tuning. This, in turn, requires materials that have tunable properties, such as graphene with tunable surface conductivity, $VO_2$ and AIST (an alloy) with temperature-dependent permittivity, magneto-optical materials with magnetically tunable permittivity, etc. Moreover, modulation can be easier by using materials that support either surface plasmon polaritons or surface phonon polaritons so that coupling between resonance surface modes of the emitter and the receiver can be controlled. Thus, authors have used DSMs which behave as metals at low frequencies and thus support surface plasmon polaritons. Notably, the permittivity of DSMs is a function of temperature (*T*), the Fermi energy ($E_f$), and the degenerate factor of 3D Dirac points (*g*) which makes it possible to tune the material properties by suitably adjusting the three parameters. Also, for a thin layer of DSM, the Fermi energy level can be controlled by simple gate voltage tuning which makes the modulation faster. In Fig. 6(c), the thermal modulator system with an emitter and a receiver is shown, that comprises a $SiO_2$ slab coated with a thin DSM film of thickness *t*, separated by a vacuum gap *d*, and kept at two different temperatures so that the heat flows from high to low temperature. The analytical form of the heat transfer coefficient (HTC), as obtained from the framework of Rytov's theory of the fluctuational electrodynamics, depends on the same three parameters like the permittivity of the DSM layer such as *T*, $E_f$, *g* as well as on *t*, and *d*. Notably, with various values of $E_f$ (ranging from 0.05 eV to 2.0 eV) and *g* (from 2 to 24), the permittivity plot of DSM exhibits metallic behavior (Re($\varepsilon_1$) <0) in the low-frequency region and thus can support surface plasmon polaritons as required for efficient modulation. Likewise, the permittivity plot of $SiO_2$ shows two resonance peaks and Re($\varepsilon_2$) <0 within the two frequency bands around the resonances, and thus readily supports surface phonon polaritons again meeting the requirements for thermal modulation. Further, to gain an insight into the physics behind the radiative heat transfer in the system, authors have rigorously studied its performance by varying parameters. Interestingly, it is observed that without the DSM layer i.e., $t = 0$, the spectral HTC exhibits resonance peaks only within the frequency bands supporting surface phonon polaritons, whereas, with an increasing *t* of the DSM layer, two new peaks arise along with the previous two peaks with an overall enhancement in spectral HTC ranging over a wide spectral band (from low to high frequency). The significant change in spectral HTC is observed with the tuning of Fermi level $E_f$ which exhibits a unique DSM-induced resonance peak shift from low to high frequency with increasing $E_f$ as shown in Fig. 6(d). Such shifting of peak into the high-frequency region is definitely beneficial for heat transfer enhancement. The origin of the DSM-induced resonance peaks is explained as the intra-film coupling (at the two DSM interfaces) resulting in low and high-frequency surface plasmon polaritons, and inter-film coupling between the low and high-frequency modes across the vacuum gap. To quantify the significant role of parameter tuning in the near field heat transfer and thermal modulation, the HTC (maximum and minimum values) and the modulation contrast $\chi$ between its maximum and minimum values are shown in Fig. 6 as a function of vacuum gap *d* with $g = 8$, $t = 20$ nm and continuously changing $E_f$ of the DSM layer which explicitly states a very high modulation contrast $\chi$ (> 40) can be achieved with a narrower *d* (< 20 nm). It is also observed that a thicker layer of DSM also exhibits high $\chi$. Such large modulation contrast readily signifies an efficient and fast thermal modulation. These findings are very encouraging for further investigations on DSM-based thermal modulator design which eventually provide significant improvement in near-field heat transfer over state-of-the-art technologies.

In this direction, in the same year, Tang *et. al.* [107] reported another novel way of active thermal modulation by relative rotation of two parallel magnetic Weyl semimetal (WSM) slabs separated by a vacuum gap *d* as shown in Fig. 6(f). We have already discussed the presence of nonreciprocal surface plasmon polaritons in WSMs in the absence of external magnetic fields. To actively control the near-field radiative heat flux, the dependence of the nonreciprocal dispersion of SPPs on the light incidence plane (characterized by the azimuthal angle $\varphi$) is the key. For a WSM slab interfaced with air, the energy dispersion curve exhibits maximum nonreciprocity (asymmetric in wavevector space) at an azimuthal incidence angle $\varphi = 0$ (upper panel of Figure 6(f)), whereas the dispersion becomes reciprocal (symmetric in wavevector space) at $\varphi = \pi/2$ (lower panel in Fig. 6(f)). The major effect of such asymmetry of WSM dispersion reflects into the HTC (scaled by the corresponding blackbody limit) when plotted with the variation of twist angle or rotation angle $\theta$ as depicted in Fig. 6(g). A maximum HTC is achieved for $0 \leq \theta < \pi/4$, whereas it becomes less dependent on the twist angle above $\pi/4$ and also varies a little with the change in gap separation d. It also shows that a minimum d gives a maximum HTC. The heat transfer control is better understood in Fig. 6(h) where the photonic transmission coefficient is plotted against the energy and the azimuthal angle $\varphi$ with twist angle $\theta = 0$ (upper panel) showing a maximum heat transfer, and with twist angle $\theta = \pi$ (lower panel) exhibiting a reduction in HTC. The red lines denote the dispersion of the surface polaritons and the blue lines are the same with $\varphi = \varphi + \theta$. This happens due to the coupling between the two surface modes; when there is no twist the modes are resonant and coupled strongly; whereas with a rotation, there is a mismatch between the two surface modes from the two WSM-air interfaces that results in weak coupling between them and thus reduces the HTC. Thus, the thermal modulation can be easily controlled by rotating any of the WSM slabs which readily provides a feasible and faster modulation technique.

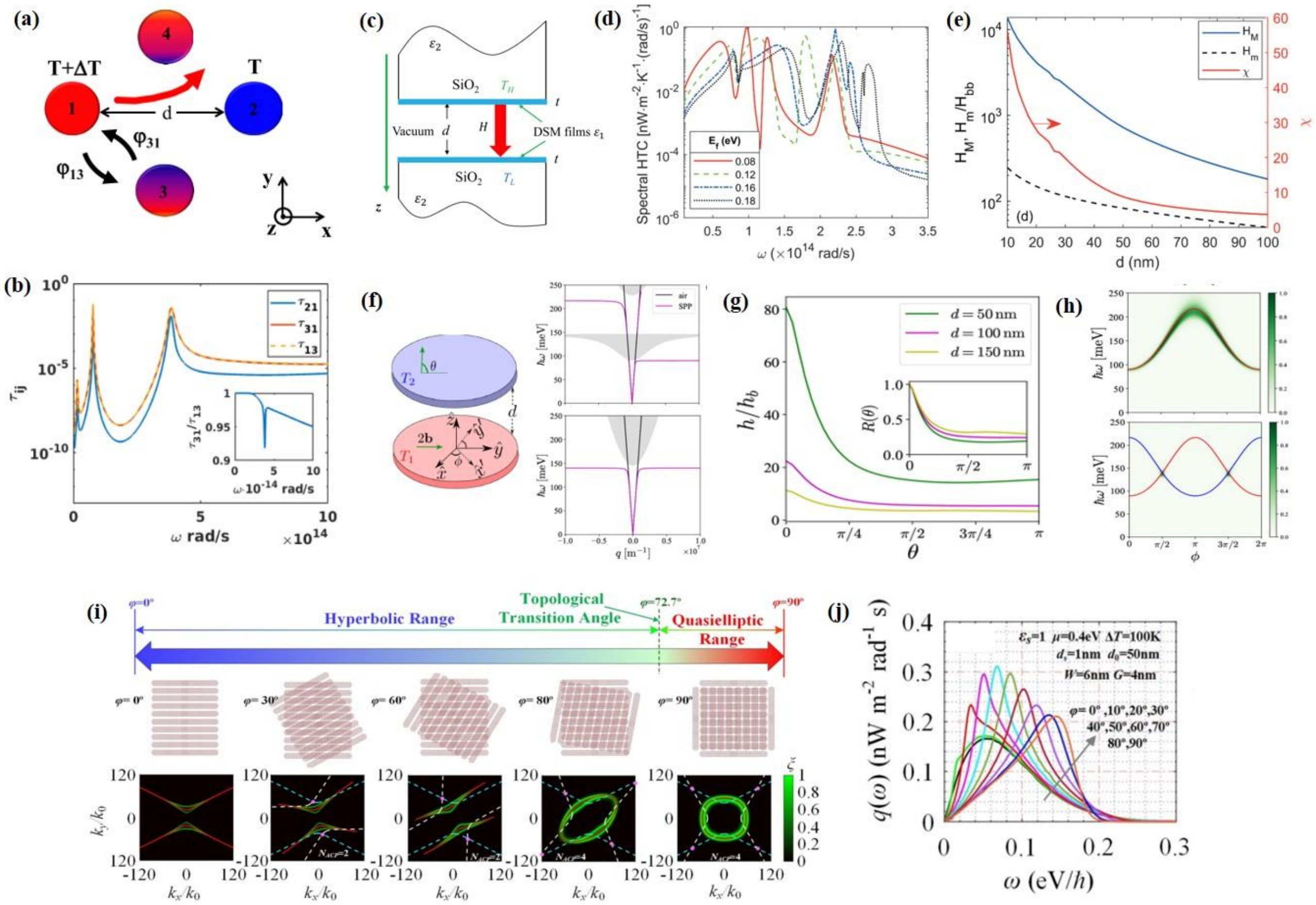

**Fig. 6**. (a) Schematics of Weyl semimetal particles. (b) Plot of transmission coefficients $\mathcal{T}_{12}$, $\mathcal{T}_{13}$, and $\mathcal{T}_{31}$ for configuration in part (a) with $d$ = 300 nm, radius $r$ = 100 nm and $T$ = 300 K, $T+\Delta T$ =310 K for four nanoparticles made of Weyl semimetals[85]. (c) Schematics of the DSM-coated near field radiative thermal modulation structure with (d) the dependence of spectral HTC over a range of Fermi level $E_f$, and (e) the dependence of maximum and minimum HTC along with modulation contrast $\chi$ over a range of gap separation $d$ [106]. (f) Schematic for near-field heat radiation between two Weyl semimetals with gap separation $d$ and twist angle $\theta$, along with dispersion curves of the surface plasmon polaritons at $\varphi$ = 0 (upper panel) and $\varphi$ = π/2 (lower panel), (g) the dependence of scaled HTC on the twist angle $\theta$ for different gap separation $d$, and (h) the photonic transmission coefficients against $\hbar\omega$ and $\varphi$ for $\theta$ = 0 (upper panel), and $\theta$ = π (lower panel), where the red lines denote the dispersion of the surface polaritons and blue lines are the same with $\varphi = \varphi + \theta$ [107]. (i) The topological transition of hyperbolic to elliptic dispersion profile in twisted-graphene-based hyperbolic topological insulator system along with the photonic transmission coefficient shown as bright green bands, and (j) the variation of spectral heat flux with different twist angle of the hyperbolic topological insulator system with $T_1$ = 310 K and $T_2$ = 300 K [108]. Reproduced with permission: ©2020, American Physical Society (a), (b); ©2021, Taylor & Francis Online (c-e); ©2021, ACS Photonics (f-h), ©2021, American Physical Society (i,j).

In 2022, Z. Yu *et. al.*[109] extended a similar concept of a twist-induced WSM-based heat flux control system to a three-body WSM for modulation and control of near-field radiative heat transfer. In this case, they considered three consecutive WSM slabs separated by vacuum gaps along the *z-axis*, and a WSM slab at one end was kept fixed while the other two WSM slabs were free to rotate in the *x-y* plane. Like the two-body counterpart, this system also exhibits enhanced radiative heat transfer owing to the nonreciprocal surface plasmon polaritons supported by the WSM slabs. Here, the WSM slab at the center along with the vacuum space at its two sides act like a cavity and contribute majorly to the increment of the HTC owing to the breaking of mirror-symmetry at specific rotational states. Authors have rigorously shown that an increased thickness of the central WSM slab increases the HTC. Moreover, the respective rotation angles $\theta$ play a crucial role for maximum surface mode couplings as is already described for the two-body counterpart; for three-body system, HTC can be estimated with various combinations of $\theta$ for all the three slabs. Both maximum and minimum HTCs are observed for specific rotation angles. In addition, the location of the central WSM slab (resulting in an asymmetric cavity) also plays a significant role in enhancement as well as adequate suppression of heat flux. Such a system can eventually be used as a thermal switch owing to its high switching ratio.

A very recent article by Y. Sheng [110] thoroughly demonstrates a material-specific study of WSM-based near-field radiative heat transfer for an ultimate optimization for heat flux modulation and enhanced nonreciprocity of surface polaritons. $Co_2MnGa$, $Co_3Se_2S_2$, $Co_3Sn_2S_2$, and $Eu_2IrO_7$ are the four materials that have recently gained attention for their unique topological properties which can be exploited to control heat transfer at the nanoscale range. Authors have provided a comparative study of the heat flux through these four materials with respect to temperature change depending on other material-specific features such as number of Weyl nodes, dielectric tensor, Fermi velocity, Drude damping rate, and chemical potential. Moreover, the influence of the gap separation of the two parallel slabs on heat flux has been reported for system design optimization purposes. Finally, an optimization algorithm based on Bayesian optimization has been provided to achieve the maximum heat flux and more nonreciprocal surface modes.

### *4.5. Graphene nanoribbons for topological radiative heat transfer*

Graphene is well-known as a 2D Dirac semimetal and has gained profound attention as a topological material. Recently, C. L. Zhou *et. al.* [108] have reported how the hybridization of surface plasmon polaritons in twisted bilayer graphene nanoribbons can lead to topological phase transition in such structures and eventually can be used to enhance the near-field radiative heat transfer. The structure comprises an emitter and a receiver kept at two different temperatures separated by a vacuum gap, and each sub-structure is made of two layers of graphene nanoribbons with a dielectric spacer in between in which one layer of graphene is free to rotate. Authors have shown that the rotation of graphene layers results in the hybridization of in-plane surface plasmon polaritons and the hyperbolic dispersion becomes elliptic by topological transition beyond a critical rotational angle. In Fig. 6(i) the effect of rotation angle on the topology of the system as well as on the photonic transmission coefficient is depicted for a specific frequency. With increased rotation angle beyond 70°, the hyperbolic dispersion lines of the two graphene layers (shown by dashed blue and white lines) are crossing each other at four points instead of two, thus making a closed loop owing to the coupling of in-plane polaritons supported by each graphene layer. Figure 6(j) explicitly shows the dependence of spectral radiative heat flux as a function of frequency on the rotation angle. A clear enhancement in the spectrum can be observed with an increasing angle for a fixed system with a 50 nm vacuum gap, a 1 nm thick dielectric spacer having permittivity equal to unity, and 0.4 eV chemical potential. Moreover, a blueshift of the spectral peak reveals that with increasing angle, the frequency of topological transition also increases, however, the amount of heat transfer appears maximum at a critical angle and then gradually decreases. Notably, the thickness of the vacuum gap and the dielectric spacer also have significant roles to play in controlling the heat transfer in such a hybridized system. It is important to note that this rotation-based hybridization of polaritons leading to topological transition is intrinsically different from the mechanical rotation-induced active heat flux control in WSM-based structures.

## 5. Time-reversal or chiral symmetry breaking

Topological phase transitions in photonic structures based on symmetry breaking and synthetic dimensions have been intensively investigated and reported [92,111,112]. It was Haldane and Raghu's phenomenal work which for the first time demonstrated the topological phase transitions in photonic geometries provided there is a broken time-reversal symmetry [4]. Photonic crystals, possessing both the time-reversal and the inversion symmetry, support the pairs of Dirac points in the Brillouin zone. In the presence of inversion symmetry, the Berry curvature $\mathbf{\Omega}_n(-\boldsymbol{k}) = \mathbf{\Omega}_n(\boldsymbol{k})$, and along with time-reversal symmetry, $\mathbf{\Omega}_n(\boldsymbol{k}) = 0$. This makes the eigenvalue problem of the system a real-symmetric type such that for all $\boldsymbol{k}$ Bloch states $\mathbf{u}_n(\boldsymbol{k}, \boldsymbol{r})$ are real, and it is possible to find the Dirac point upon parametric tuning. Breaking the inversion symmetry only makes the Berry curvature nonzero, however, to attain the nontrivial topology, breaking of time-reversal symmetry is necessary which opens a gap at the Dirac point with nonzero Chern numbers. On the other hand, chirality reversal symmetry is specific to certain materials like bianisotropic materials owing to the presence of photon spin angular momentum. Both types of symmetry can be broken by using an external field. Such time-reversal and also chirality-reversal symmetry breaking with an effective non-zero pseudo-Berry curvature generates chiral edge states or Floquet states [4,7,113,114]. Besides, the realization of nontrivial topology through synthetic dimensions has shown robust wave streams across a photonic network [115,116]. Very recently, researchers have shown that the application of time-reversal or chiral symmetry breaking eventually enables the enhancement of near-field thermal radiation transfer.

Regardless of topological structure, nonreciprocal many-body systems with periodic external modulation show Berry phase-induced pumping and additional effect on heat transfer [117,118]. In addition to the dynamic phase, the geometric Berry phase can act as a heat pump providing an additional heat flux across junctions even without on average no thermal gradient. This phase-induced additional heat flux shows huge differences in relaxation time of hot bodies before one cycle of the modulation ends.

In 2019, Khan et al. [119] proposed a theory to generate elliptically polarised thermal emission from a simple flat planar configuration of a topological insulator as an emitter without requiring any complex structured surfaces. They have shown that a semi-infinite slab of topological insulator with a thin magnetic film coating followed by an air interface, as depicted in Fig. 7(a), readily breaks the time-reversal symmetry of the system which in turn results in topological edge states. However, for the emission of thermal radiation, it is necessary to have a loss in the system along with the magnetoelectric coupling, owing to the distinguishable absorption present in topological insulators [120,121], which ultimately leads to a nonzero average spin angular momentum of photons with elliptically polarized eigenstates. At equilibrium, topological insulators contribute to thermal radiation via their surface edge states through inelastic backscattering with phonons. The spin-momentum locking of scattered electrons creates a non-zero correlation among fluctuating spin currents, creating thermal radiation with some degree of circular polarisation; consequently, the average spin angular momentum is not zero. Authors have shown the variations of the normal component of spin angular momentum per photon denoted by $S_n$ with the magnetoelectric coupling $\kappa$ and loss to establish that both are important to realize a nonzero $S_n$. Moreover, the dependency of $S_n$ on the emission angle, $\vartheta$ implies that a slab of topological insulator radiates an equal amount of $S_n$ for all given angles, unlike a structured interface[122] rendering them acceptable candidates for structured light generation from thermal sources that can be applied to the detection and management of radiation in thermal emitters.

### *5.1. Radiative heat transfer of topological slabs in time-reversal symmetry breaking fields*

Near-field thermal radiation always exhibits an improved spatial coherence compared to far-field thermal emission which induces many interesting phenomena to explore. Such spatial coherence is also beneficial for the enhancement of thermal radiation. In 2013, Xiao et al. [123] reported that the thermal coherence properties of a topological insulator slab with gapped surface states are much improved when compared with radiation from a dielectric slab. The authors considered a system where a topological insulator slab of thickness $h$ is kept in a magnetic field **B** as shown in Fig. 7(b). The presence of the magnetic field **B** opens a gap in the surface of the topological insulator by breaking the time-reversal symmetry, and also the strength of **B** is directly proportional to the surface gap $\Delta$. It is shown that there are three regions of operation based on the relation between the surface gap $\Delta$ and the radiative photon energy $\hbar\omega$ in a topological insulator slab with gapped surface states that contribute to the enhancement of spatial coherence of thermal radiation. In the interband transition region, where $\Delta << \hbar\omega$, the coupling between the surface states and the waveguide modes in the bulk of the topological insulator are weaker which does not contribute to any improvement in the spatial coherence. With $\Delta \geq \hbar\omega$, one finds an operation regime where surface states get coupled to the waveguide modes and also start inducing a resonance to the transmission channel leading to TM modes in vacuum. This phase is quite interesting as one may observe changes or effects on topological insulator slab thickness $h$, components of cross-spectral density tensor $\boldsymbol{W}_{xx}$ (or $\boldsymbol{W}_{yy}$), and the coherence length, all based on an increasing Δ. To note, $h$ decreases as Δ increases, whereas spatial coherence and coherence length are both found to be enhanced with an increasing Δ. When $\Delta >> \hbar\omega$, $h$ reaches a minimum and becomes independent of Δ change. This regime exhibits a half-integer quantum Hall effect in the topological insulator slab where the surface states acquire half-integer quantized surface Hall conductivity (HIQSHC), and a strong coupling between the surface states and waveguide modes is observed with highly enhanced spatial coherence and a longer coherence length. The coupling of gapped surface states and waveguide modes in the topological insulator slab results in a novel resonance that modifies the ratio of transverse magnetic (TM) and transverse electric (TE) components in the radiation field. As $\boldsymbol{W}_{xx}$ increases with increasing Δ, components of TM modes are higher in the radiation which also shows better coherence than TE modes thus the overall spatial coherence gets enhanced. Moreover, the authors discussed the change in the ratio of the TM and TE modes of the radiative field and its significant impact on the coherence length, $l_{xx}$, as a function of the surface gap, Δ, at different frequency bands. In the inter-band transition regime, the coherence length is inconsiderable and does not vary significantly with the surface gap; in the crossover regime, $l_{xx}$ sensitively increases with the surface gap, whereas in the HIQSHC regime, $l_{xx}$ approaches its maximum value and plateaus with increasing surface gap. In this regime, $l_{xx}$ is virtually 10 times larger than the coherence length of $\boldsymbol{W}_{xx}$ for the dielectric slab. It is also observed that in the HIQSHC regime, the gapped surface state could significantly enhance the thermal radiation energy density by up to 7500 times more than that of the blackbody radiation energy at a given frequency.

*5.2. Radiative heat transport in closed orbits*

In recent years, the substantial enhancement of radiative heat transfer in the presence of a temperature gradient (non-equilibrium) in non-reciprocal systems has been demonstrated that includes thermal rectification [124], thermal Hall Effect [82], and tailored radiative heat transfer [125]. Here, we review the work of M. G. Silveirinha [78] that specifically demonstrated the impact of the topological edge state of light on radiative heat transfer in a closed cavity at thermal equilibrium and observed that a persistent flow of electromagnetic momentum exists within the cavity even at zero temperature. The geometry under consideration (as depicted in Figs. 7(c-d) comprises a parallel plate (metallic) waveguide filled with a gyrotropic material under the influence of a static magnetic field and directed to the negative $y$-axis. The topological property of the bulk gyrotropic material and the behavior of the topological interface states in radiative heat transfer are studied for a specific high-frequency bandgap that lies within a frequency range determined by the plasma frequency and the cyclotron frequency related to the applied magnetic field. Specifically, the author has shown that there is a single propagating mode irrespective of the propagation direction in the plane of propagation (Fig. 7(c)). The mode dispersion for propagation along $x$ and $y$ directions within the frequency interval (shown by the grey dashed horizontal lines) is depicted in Fig. 7(e). The dispersions for $x$ and $y$ directions are shown by blue and green lines which are nearly coincident. The dispersion of the topological edge mode is shown by the black line with negative $k$-values when the top plate is removed. Interestingly, the electric field is found to be dominant along the z-direction which is consistent with the boundary conditions by the metallic plates but sharply contrasting with the electric field polarization for the bulk gyrotropic. Moreover, the energy flow is also direction dependent such that for a +x propagation direction, energy tends to flow near the top metal plate, whereas, along -x direction, it mostly flows near the bottom plate i.e., an asymmetric energy flow (Fig. 7(f)); for propagation along y, the energy flow is symmetric with respect to the waveguide center (Fig. 7(g)). Such spatially asymmetric energy flow has a topological origin which shows that there are topological states at the two metal-gyrotropic interfaces, and the electric field distribution of the entire waveguide is in sync with that of the topological edge state at a single interface. Furthermore, the role of topological modes in radiation heat transfer is demonstrated by using the Poynting vector spectral density. The expectation value of the spectral density of the Poynting vector is nonvanishing, even at thermal equilibrium, owing to the topological nature of the gyrotropic material. The existence of topological edge states results in the annihilation of the Poynting vector close to the waveguide center and enhancement near the metallic wall where a unidirectional current exists. Moreover, as the circulation of heat flow in closed orbits occurs in the gyrotropic waveguide, the electromagnetic angular momentum is nontrivial while the total electromagnetic momentum of the system is zero, which is attributed to the system even at zero temperature where the fluctuations are purely quantum in nature. The unidirectional nature of thermal fluctuations in such systems can be further explored for unidirectional heat transport.

### *5.3. Thermal spin photonics*

Following the previous work on the effects of topology on radiative heat transport in closed cavities, Khandekar *et. al.* [81] focused on the generalization of spin-momentum locking of light to thermally excited waves and experimental detection of persistent thermal current. For this, they started with preparing a more generalized theoretical framework for fluctuational [74,82,126–128] and quantum [129–133] electrodynamic effects using generic bi-anisotropic materials. Notably, authors have presented a persistent thermal photon spin (PTPS) and a persistent planar heat current (PPHC) in the near field thermal radiation of the InSb slab (Fig. 7(h)) which is a non-reciprocal material owing to the antisymmetric permittivity tensor in presence of a magnetic field and also supports thermally-excited spin-momentum locked surface plasmon-polaritons. It is found that the spin-momentum locked polaritons of InSb also exhibit an asymmetric spin magnetic moment resulting from the asymmetry in the spin angular momentum and heat flux carried by thermally excited polaritons. A detailed analysis of the frequency spectrum of energy density, spin angular momentum density, and Poynting flux at a distance of 1 μm from the InSb slab surface at thermal equilibrium is reported. From the analysis of spin-momentum locked polaritons, the contribution of thermally excited waves is schematically shown in Fig. 7(h) which shows the waves having the spin component parallel to applied magnetic field and the waves having their spin component anti-parallel to applied magnetic field. The positive component is red-shifted, whereas the negative component is blue-shifted, thus broadening the total energy density. Interestingly, when no magnetic field is present, the asymmetric contributions of the positive and negative parallel components of the wave vector result in non-zero spin angular momentum and nonzero Poynting flux leading to PTPS and PPHC despite the thermal equilibrium where both geometry and vacuum are maintained at 300 K. The detailed study shows a higher contribution of the electric-type spin to the PTPS than the magnetic-type spin which also holds universally for any gyroelectric type nonreciprocal material. The dispersion curves of surface polaritons making three different angles with the applied magnetic field are shown in Fig. 7(i). The polaritons with positive spin components along the applied magnetic field are red-shifted, whereas those with negative spin components are blue-shifted. Dispersion for $\varphi = 0$ is the same for all angles without the magnetic field or with the magnetic field perpendicular to the surface. The authors also proposed an experimental setup that could be employed to derive direct evidence of the existence of a persistent planar heat current. The proposed setup is shown in Fig. 7(j). The proposed system consists of a planar InSb slab covered by an aqueous medium containing suspended non-magnetic microparticles. The system is in thermal equilibrium at room temperature, and Brownian movement occurs around the mean positions of microparticles. When a magnetic field is applied, the microparticles gain additional translational and rotational properties between spin-momentum locking and radiative heat transfer for exploring and understanding new ways of achieving unidirectional heat transfer at the nano-scale level. owing to the PTPS and persistent planar heat current. This work shows the importance of the connection

## 6. Other 2D or 3D topological insulator

In this section, we specifically review the influence of other 2D or 3D topological insulators on near-field radiative heat transfer. They show unique dispersive characteristics of the surface plasmon polaritons, leading to great attention. Notably, a topological insulator is different from a conventional insulator owing to its altered bulk topology via bandgap closing and re-opening. This leads to a conductive surface state and insulating bulk according to the bulk-edge correspondence as described in the topological band theory under section I. These topological insulators have recently been reported employing materials like $Bi_2Se_3$ which is found to exhibit pronounced nontrivial surface states and dominate the radiative heat transfer in the nanoscale. $Bi_2Se_3$ is best known as a topological insulator rather than as a typical Dirac semimetal like graphene. Such topological insulators exhibit conductive surface states with insulating bulk both in 2D and 3D.

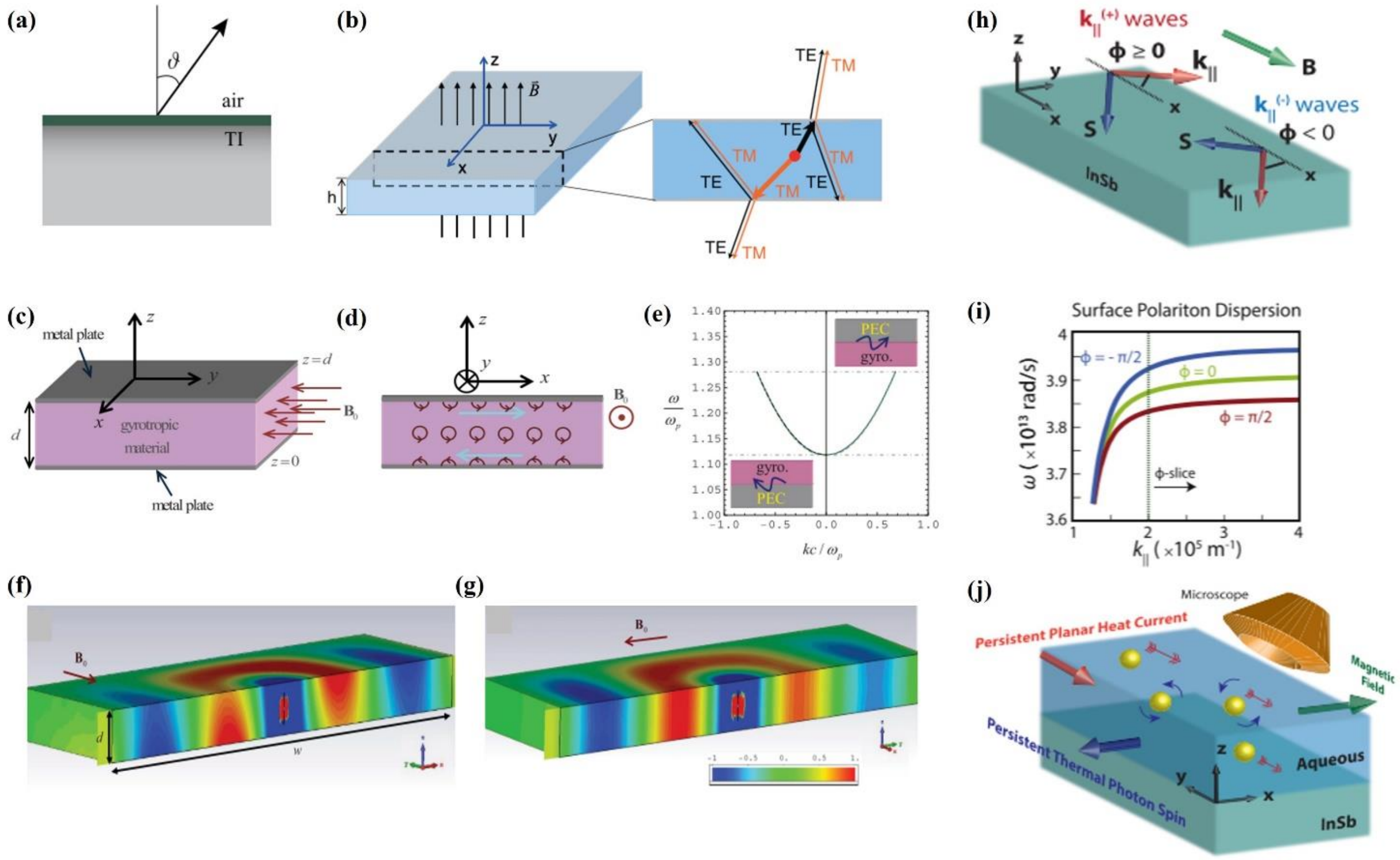

**Fig. 7**. (a) Schematics of topological insulator coated with thin magnetic film as thermal emitter [119]. (b) (Left) Schematics of topological insulator slab considered in this work; (Right) scattered thermal radiation fields at topological insulator surfaces [123]. Schematics of geometry considered in [78]. (c) Closed cavity made of parallel plane waveguide filled with gyrotropic material. (d) Schematics of electron cyclotron orbits in gyrotropic material; field radiated by a vertical electric dipole in waveguide. (e) Mode dispersion for propagation along *x* and *y* directions. (f) View of *xz* plane. (g) View of *yz* plane. Density plots demonstrate the time slot of the radiated electric field in the *z-direction*. (h) Schematics of geometry are considered in this work [81]. (i) The dispersion plot of surface polaritons at different angles $\varphi$ made by polaritons with the applied magnetic field B. (j) A planar slab of InSb coated with an aqueous medium having suspended micro-scale non-magnetic particles at room temperature. Reproduced with permission: ©2019, American Physical Society (a); ©2013, American Physical Society (b); ©2017, American Physical Society (c-g); ©2019, IOP Science (h-j).

R. Liu *et. al.* [134] have recently investigated the nontrivial surface plasmon polaritons for near-field heat transfer in a system of two monolayered $Bi_2Se_3$ sheets kept at different temperatures and separated by a vacuum gap. It is observed that coupling between surface plasmon polaritons supported by each vacuum-$Bi_2Se_3$ interface leads to strong tunneling of thermally excited photons via surface polaritons, thereby increasing the heat flux. Figure 8(a) schematically depicts the topological insulator system, whereas Figure 8(b) shows the photonic transmission coefficient which clearly shows the strong bright branches of surface states split into symmetric and antisymmetric modes as a result of polariton coupling. Notably, the heat flux increases as the vacuum gap becomes narrower (<10 nm).

Moreover, a red shift of the peak in the spectral heat flux results from an increasing vacuum gap. Authors have also reported the effect of the Fermi level on heat transfer. As can be observed in Fig. 8(c), there is an optimum Fermi level (0.1 eV) for which strong and continuous surface states appear that mostly contribute to the enhancement of heat transfer, below and above which the heat transfer will be significantly reduced. In addition, a practical topological insulator system with a non-polar and non-dispersive substrate has been considered instead of a vacuum and a drastic reduction in the heat flux has been observed which readily dictates that to implement such a 2D topological insulator system in practice, the choice of substrate material is of utter importance due to the pronounced interference effect between surface modes and substrate modes.

In 2022, H. Wu *et. al.* [135] have taken a step forward in this direction, and thoroughly reported the potential of near-field heat transfer in a 3D topological insulator system in the presence of both the nontrivial surface states and the insulating bulk modes of the finite thickness $Bi_2Se_3$ substructures, in contrast to the 2D $Bi_2Se_3$ monolayer system reviewed above. The 3D topological insulator system considered is shown in Fig. 8(d) which comprises two identical sub-structures which are $Bi_2Se_3$ films kept at two different temperatures and separated by a vacuum gap. In this way, each topological insulator film has two topologically conductive states at the two vacuum-Bi2Se3 interfaces with insulating bulk states in between. Figure 8(e) shows the energy transmission coefficient (left upper), dispersion (middle upper), and spectral heat flux (right upper) of the 3D topological insulator. Here, one would observe radiative heat transfer through the formation of hybrid modes by coupling the Dirac plasmons at the interfaces with the phonon polaritons of the bulk, resulting in a hyperbolic plasmon-phonon polariton (HPPP) within the hyperbolic bands, and surface plasmon-phonon polaritons (SPPP) outside the hyperbolic band region. In the low-frequency region $\omega < 64$ cm$^{-1}$, there is a small contribution from coupled SPPP to energy transmission resulting in a small amount of heat flux. In the frequency range 64 cm$^{-1}$ $< \omega <$ 163 cm$^{-1}$, the interactions of plasmons and phonon polaritons are more complex, and different regions based on mode interactions are formed which has a significant effect on the spectral heat flux. Here, the contributions from hybrid modes are absent owing to the non-local effect of the surface states. The sharp peak in the spectral heat flux at $\omega \approx 142.5$ cm$^{-1}$ is due to the presence of surface phonon polaritons only. Another intense spectral peak at $\omega \approx$ 161 cm$^{-1}$ is the contribution of the flatter dispersion of the HPPP within the type-I hyperbolic band. Also, in the high-frequency range, SPPP are present due to coupling of top and bottom surface plasmons which have a maximum contribution in the type-I hyperbolic band resulting in a sharp peak at $\omega \approx 165$ cm$^{-1}$ and decaying rapidly above this. It is also observed that increasing the vacuum gap decreases the heat flux, however, an increase in the film thickness readily increases the heat flux for a fixed chemical potential. Moreover, the chemical potential has a significant role in controlling the heat flux in the high-frequency region with a broad spectrum through the high-frequency SPPP of surface states. Thus, more active control of radiative heat transfer can be achieved in 3D topological insulators owing to their unique surface and bulk mode interactions. Very recently, based on such 3D topological insulators, a three-body system has been proposed by S. Fu *et. al.* [136] showing an enhanced radiative heat transfer for both doped and undoped cases compared to a two-body system depending on optimized vacuum gaps and thickness of the intermediate topological insulator film.

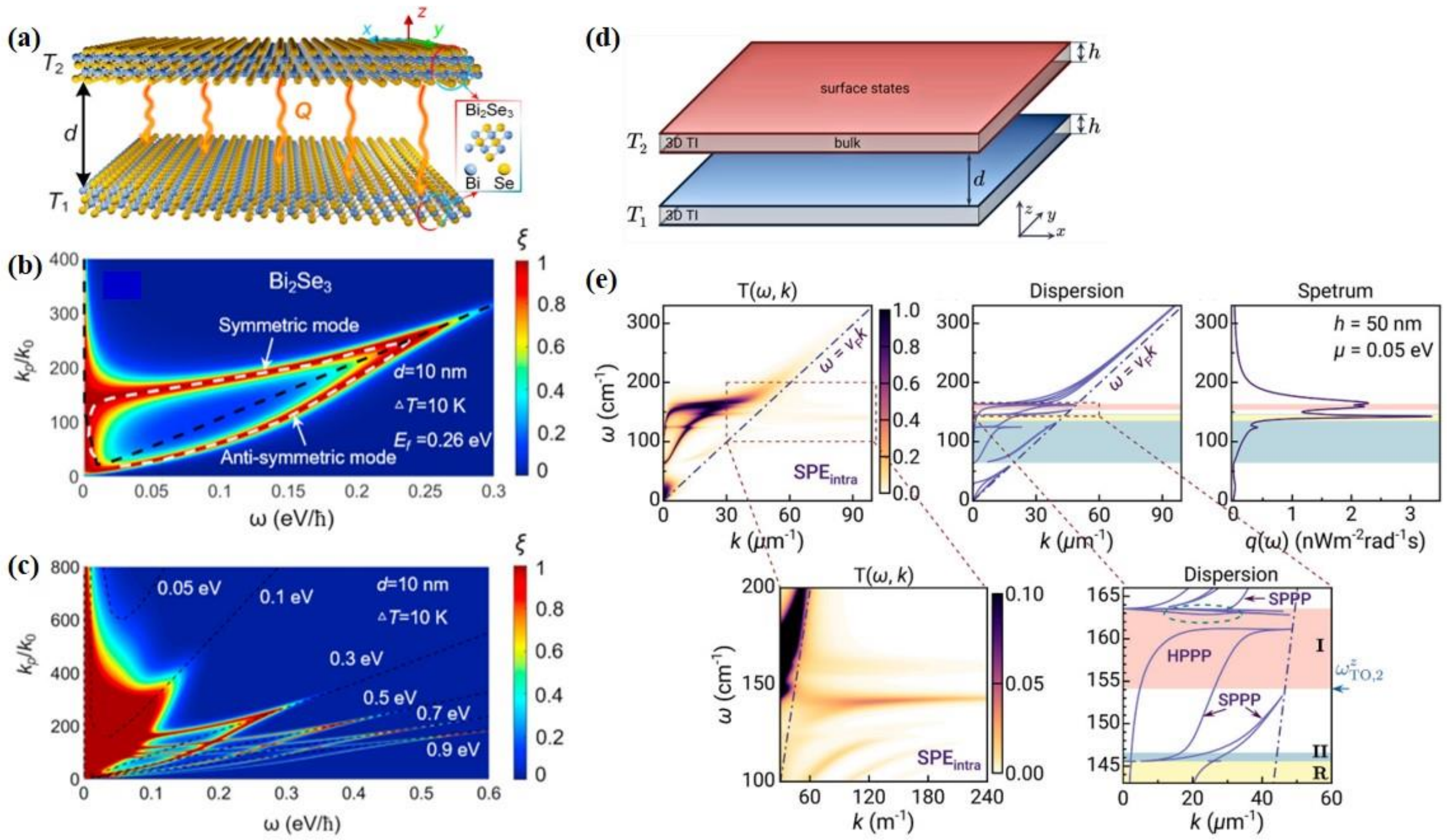

**Fig. 8**. (a) The schematic representation of the 2D $Bi_2Se_3$-based monolayered topological insulator system kept at two different temperatures $T_1$ (300 K) and $T_2$ (310 K), and the photonic transmission coefficient of the system at a 10K temperature difference with 10 nm vacuum gap at (b) 0.26 eV Fermi level exhibiting the splitting of modes, and (c) 0.05, 0.1, 0.3, 0.5, 0.7, and 0.9 eV Fermi level to show its dependency on Fermi energy; the black dashed lines are the dispersion profile of the system [134]. (d) The schematic of the 3D $Bi_2Se_3$-based topological insulator system with $T_1$ = 290 K and $T_2$ = 300 K, (e) the energy transmission coefficient T($\omega$,$k$), dispersion, and spectral heat flux $q(\omega)$ of the 3D topological insulator system shown in the upper panel; the lower panel shows the close view of the portions of the curves inside the dotted box [135]. Reproduced with permission: ©2021, Science Direct (a-c); ©2022, Elsevier (d,e).

## 7. Conclusions and Outlook

In this review, we aimed to present the significant contribution of topological photonics in the domain of near-field thermal radiation which essentially opens up new platforms for futuristic topology-protected thermal devices for various applications such as sensing, routing or thermal switching. Some of the fundamental concepts in the two fields are covered. First, we have shed light on the fundamentals of topological photonics and photonic materials that exhibit nontrivial topology and edge states of light. Then, the fundamentals of radiative heat transfer in the near field and thermal radiation in the super-Planckian regime of heat transfer are discussed in detail; this regime surpasses the classical predictions of radiative heat transfer given by Planck. We also reviewed the wide range of topological photonic systems reported to date for thermal applications. These systems are insulators in bulk and conductors on the surface, enabling the unidirectional flow of energy because of the existence of their edge states. Owing to the strong localization at the surface and protected by the bulk topology, the topological edge states are eventually exotic and robust for enhanced near-field thermal radiations compared to other conventional surface states. Therefore, it is high time to explore this domain inquisitively for future developments.

Subsequently, the review is categorized into five domains based on the type of systems, covering thermal transport in SSH-based topological systems, quasi-periodic nanoparticle chains exhibiting topological features, systems with Dirac and Weyl semimetals, systems in which time-reversal or chiral symmetry breaking is achieved, and other 2D and 3D topological insulator-based models.

The review of SSH-based systems highlighted studies on radiative heat flux through topological SSH chains of nano-particles. In these systems, we reviewed studies that investigated the spectral power received by SSH chains of nano-particles in topologically trivial and non-trivial regions. In these systems, the heat flux of the longitudinal modes is mainly due to the propagating modes in the topologically trivial region, and the heat flux of transverse modes is due to the band modes. However, in the topologically non-trivial case, the heat flux of both the transverse and longitudinal modes is due to the edge modes. Furthermore, despite the expectations arising from the theory of heat transfer at the nano-scale, the long-range coupling between the first and last nano-particles can be achieved in the SSH chains of nano-particles due to retardation. This behavior shows the effect of the edge mode heat flux for any given length of nano-particles. The study of SSH-based systems further reveals that the edge and corner states in the topological non-trivial region can increase the energy density in the proximity of the edges and corners in 1D and 2D SSH-based plasmonic or phonon–polaritonic nano-particles. In addition, the temperature dependence of the permittivity of SSH chains (i.e., InSb chain) enables tuning the resonance frequency of TPPs using temperature.

The inclusion of quasiperiodic models into the array of nanoparticles to achieve nontrivial surface states has been reviewed as the second category of topological systems that has gained attention recently for near-field radiative heat transfer. Such quasiperiodic models have been investigated in topological photonics for quite some time, however, their use for thermal devices has surged lately. The use of the well-known AAH and Fibonacci models in nanoparticle chains results in nontrivial phonon polariton states since it inherits the topological property of two-dimensional integer quantum Hall systems. Moreover, the radiative heat transfer can be intrinsically controlled or manipulated to a good extent owing to the modulation phase present in the model. However, such topological chains require more in-depth investigations in this direction, mainly from theory to experiment for their practical implementations.

The third category of topological models reviewed for application to radiative heat transfer was that of Dirac and Weyl semimetals. Weyl semimetals belong to 3D gapless topological models; their non-trivial topology enables exceptional optical and electronic characteristics. In the context of radiative heat transfer, these characteristics can be used to design thermal routers, non-reciprocal thermal emitters, and compact optical isolators. Furthermore, achieving time-reversal symmetry breaking through magnetic Weyl semimetals without the necessity of applying an external magnetic field was found to be possible. The axion electrodynamics in topological magnetic Weyl semimetals was also examined. It allows for the creation of unique non-reciprocal thermal emitters that can virtually completely violate Kirchhoff's law. Also, active modulation/control of heat flux using a group of materials exhibiting Weyl nodes as well as exceptional topological properties have gained a huge interest of researchers, and optimization of Weyl semimetal-based systems for heat flow enhancement or suppression should lead to the next-level nonreciprocal thermal emitters and routers for nano-scale energy harvesting and heat transport.

The fourth category considered pertains to systems that achieve time-reversal or chiral symmetry breaking. The review found that radiative heat transfer through topological slabs capable of achieving time-reversal breaking could be designed for structured light generation from thermal sources. A potential application of the foregoing is the detection and management of radiation in thermal emitters. Furthermore, the thermal coherence properties of a topological insulator slab indicate that when the surface gap exceeds the radiative photon energy, the strong coupling between the surface states and waveguide modes can significantly change the coherence characteristics of thermal radiation. The connection between spin-momentum locking and radiative heat transfer was also examined to better comprehend possible new ways of achieving unidirectional heat transfer at the nano-scale level.

The fifth category discussed the potential of higher dimensional topological insulators in near-field radiative heat transfer. We have observed that a system with two graphene nanoribbons can participate in topological phase transition by giving a rotation to one of the monolayers resulting in enhanced nontrivial surface states, thereby leading to enhanced radiative heat flux. Moreover, the addition of a certain amount of thickness in the 2D topological insulator structures gives rise to exclusive topological features owing to the presence of both the surface and bulk states and hybrid modes from their coupling. Such hybrid modes as well as surface plasmons and phonon polaritons in such 3D topological insulators intrinsically modulate the heat flux depending on the frequency of operation, and structural parameters. This domain is utterly new and intriguing and thus requires more investigations theoretically as well as experimentally.

Even though there have been many theoretical studies available, experimental realizations are still in the early stage, mainly owing to the difficulties and challenges in the near-field radiative transfer setup [137]. The setup requires precise control and measurement of sub-wavelength gaps between the source and detector in the range down the order of tens of nms. If the gap distance does not need to be changed, under-etching processes in the semiconductor fabrication process may be considered. If the gap distance needs to be changed, the piezo-driven translational stages are necessary, but optical distance measures are also required to prevent contact damage. Moreover, the choice of detectors for thermal radiation also requires high-resolution thermal imaging devices or well-calibrated single-channel detectors, which ensures the seamless measurement of spectra. In addition, the whole system should be located in a thermally shielded chamber where external interference from external factors such as ambient temperature fluctuations and electromagnetic radiation noise are sufficiently minimized.

Towards this, L. Cui et. al. [138] have demonstrated a thorough experimental technique to measure heat transfer below 10 nm to Angstrom-sized gaps between a heated Au-substrate and custom-fabricated Au-coated scanning thermal microscopy probe (SThM) in an ultra-high vacuum chamber. Their work specifically emphasizes the step-by-step cleaning processes that are extremely important, so no other heat transfer channel gets created due to impurities and contaminations at such a small gap. Such contaminations at the nanometer scale or less than that are evident and also detrimental to near-field radiative heat transfer. To reduce the surface contamination effect, authors used both the ex-situ (oxygen plasma-cleaning process) and the in-situ (controlled crashing) cleaning process which eventually rectified the transmittance. Furthermore, H. Salihoglu et. al. [139] have demonstrated radiative heat transfer within a 10 nm gap distance between two bulk, rigid Quartz plates. Authors have used an interference spatial phase matching process for the gap measurement, and microfabricated sensors for the measurement of temperature variation due to radiative heat transfer while the total arrangement was kept in ultra-low particulate air filters. Authors have claimed that their experimental technique provides a better way to measure and control gap distances within the sub-nanometer range and may also be employed for metasurfaces. Such recent experimental reports are truly encouraging for further developments in this direction.

Besides the developments in the experimental setup, it is also important to choose the materials in the interacting surfaces with well-defined thermal and optical properties. Notably, the recent experiments are performed mostly with bulk, rigid samples, whereas topological materials need special attention owing to their distinct features. Recently, two experiments on near-field radiative heat transfer between graphene-covered SU8 heterostructures have been reported, which eventually encourages such experiments with topological materials [140,141]. Also, the quality of the topological materials is an important factor since the features of topological light states are strongly bound to their geometric parameters. Thus, an accurate fabrication of such materials dictates the success of the experiment. Recently, 3D micro-printing devices are commercially available to fabricate photonic crystal and waveguide-based topological structures [142].

Table 1 summarises the research papers reviewed in this work, categorizing them according to the year of publication, subject, selected material compositions, and structures considered.

Lastly, it is necessary to say that the interplay of topological photonics and radiative heat transfer is not limited to the reviewed models or the near-field regime. It is indeed possible to extend these studies to cover the far-field regime of radiative heat transfer [143] that may be applied to solar thermo-photovoltaic absorbers, optical bandpass filters, and sub-diffraction thin-film lenses. Upon the development of more topological models and lattice structures, which include (but are not limited to) Rice–Mele, Hatano–Nelson, Aubry–André–Harper, and Kitaev models with Kagome or Lieb lattices, radiative heat transfer with topologically protected robustness can be further investigated and harnessed considering various combinations and geometries.

**Table 1.** Summary of recent developments in topological photonic studies from the perspective of near-field radiative heat transfer.

| References (Year) | Subject of Study | Topological model | Material | Structure | Gap separation |
|---|---|---|---|---|---|
| Fu et. al.[136] (2023) | Heat flux control using 3D Topological insulators | 3D Topological insulator slabs | $Bi_2Se_3$ | Three 3D topological insulator slabs | Two vacuum gaps each of 50 nm |
| Y. Sheng[110] (2023) | Optimization of the Weyl semimetal system | Weyl semimetals | $Co_2MnGa$, $Co_3Se_2S_2$, $Co_3Sn_2S_2$, and $Eu_2IrO_7$ | Two Weyl semimetal discs | 20 nm |
| Wang et. al.[99] (2023) | Heat flux control in quasiperiodic 1D topological chains | IAAF | SiC | Hybrid nanoparticle array | Lattice constant = 0.6 µm, 100 NPs |
| Wang et. al.[98] (2023) | Heat flux control in quasiperiodic 1D topological chains | AAH | SiC | AAH model-based nanoparticle array | Lattice constant = 0.6 µm, 100 NPs |
| Wu et. al.[135] (2022) | Study of hybrid surface states in 3D Topological insulators | 3D Topological insulator slabs | $Bi_2Se_3$ | Two 3D topological insulator slabs | 10 nm |
| Yu et. al.[109] (2022) | Active heat flux modulation in three body systems by mechanical rotation | Weyl Semimetals | - | Three Weyl semimetal slabs | 100 nm (including two vacuum gaps and one thin layer) |
| Liu et. al.[134] (2021) | Heat flux control using 2D Topological insulators | 2D Topological insulator slabs | $Bi_2Se_3$ | Two 2D monolayer topological insulator films | 10 nm |
| Zhou et. al.[108] (2021) | Twist-induced topological transition in graphene nanoribbons | Graphene nanoribbons | Graphene | Two graphene nanoribbon-based systems with dielectric spacers | 50 nm |
| Tang et. al.[107] (2021) | Active heat flux modulation by mechanical rotation | Weyl semimetals | - | Two Weyl semimetal discs | 100 nm |
| Xu et. al.[106] (2021) | Active heat flux modulation using Dirac semimetals | Dirac semimetals | - | Dirac semimetal layer on $SiO_2$ slab | 20 nm |
| Ott et al.[90] (2021) | 1D and 2D topological chain | SSH | InSb | Chain of nano-particles | Lattice constant = 1µm, 10 NPs |
| Ott et al.[89] (2020) | 1D topological chain | SSH | InSb | Chain of nano-particles | Lattice constant = 1 µm, NPs varies from 20 to 500 |
| Ott et al.[85] (2020) | Anomalous thermal Hall effect | Weyl particles | InSb | Network of particles | 300 nm |
| Zhao et al.[102] (2020) | Axion field-induced thermal radiation | Magnetic Weyl Semimetals | - | Nano-spheres | - |
| Guo et al.[104] (2020) | Thermal router | Tuneable Magnetic Weyl Semimetals | - | Nano-spheres | 320 nm (centre to centre gap) |
| Khan et al.[119] (2019) | 2D topological insulator | A slab of topological insulator | - | Slab | - |
| Tang et al.[86] (2019) | 1D topological chain | SSH | Carbon | Chain of nano-particles | 2 to 6 nm for maximum heat flux |
| Wang et al.[92] (2019) | Topological plasmon for terahertz radiation sensing | SSH | InSb | Chain of micro-particles | Lattice constant = 10 μm, 100 NPs |
| Khandekar et al.[81] (2019) | Persistent thermal current | Bianisotropic Planar Slab | InSb | Slab | 1 μm above from the surface of InSb |
| Silveirinha[78] (2017) | Gyrotropic waveguide | Topological angular momentum | Gyrotropic material | Slab | $d = 0.5c/\omega_p$, c=light speed, $\omega_p$=Plasma frequency |
| Rodriguez-López et al.[101] (2015) | Two Dirac material sheets | 2D Chern insulators | Graphene | Parallel sheets | 10 nm |
| Xiao et al.[123] (2013) | Time-reversal symmetry breaking | topological insulator | - | Slab | $\lambda_{max}/20$, $\lambda_{max}$ = wavelength of the maximum radiation frequency |

## Author Contributions

Dr. Azadeh Didari-Bader, Dr. Seonyeong Kim, and Heejin Choi contributed equally to the manuscript.

## Declaration of Competing Interest

The authors declare that they have no known competing financial interests or personal relationships that could have appeared to influence the work reported in this paper.

## Funding

This work was supported by the National Research Foundation of Korea (NRF) through Grant Nos. NRF-2019K1A3A1A14064815, 2020R1I1A3071811, 2021R1A2C2010592, and 2022M3H3A1085772. Prof. Chang-Won Lee acknowledges the support received from Hanbat National University in 2022; and Dr. Heejeong Jeong acknowledges the University of Malaya Impact Oriented Interdisciplinary Research Grant (IIRG001-19FNW) and the support of Air Force Office of Scientific Research (FA2386-20-1-4068).